\documentclass[12pt]{article}
\pdfoutput=1

\usepackage[applemac]{inputenc}
\usepackage{jheppub}
\usepackage{ifpdf}
\usepackage{amssymb}
\usepackage{amsfonts}
\usepackage{amsmath}
\usepackage{graphicx,psfrag, subfigure}
\usepackage{dsfont}
\usepackage{comment} 
\usepackage{braket}
\usepackage{placeins}

\newcommand\be{\begin{equation}}
\newcommand\ee{\end{equation}}
\newcommand{\bea}{\begin{eqnarray}}
\newcommand{\eea}{\end{eqnarray}}
\newcommand\dd{{\rm d}}

\newcommand{\g}{\ensuremath{g_{a\gamma}}}

\newcommand{\mug}{\,\mu\hbox{G}}

\newcommand{\axion}{\Ket{a}}

\newcommand{\s}{\rm\; s}
\newcommand{\ks}{\rm\; ks}
\newcommand{\keV}{\rm\; keV}
\newcommand{\cm}{\rm\; cm}
\newcommand{\erg}{\rm\; erg}
\newcommand{\ergpcmsqps}{\hbox{$\erg\cm^{-2}\s^{-1}\,$}}
\newcommand{\pcmsq}{\hbox{$\cm^{-2}\,$}}

\numberwithin{equation}{section}
\begin{document}
\begin{titlepage}

\setcounter{page}{1} \baselineskip=15.5pt \thispagestyle{empty}

\bigskip\
\begin{center}
{\Large \bf  A New Bound on Axion-Like Particles}
\vskip 5pt

\vskip 15pt
\end{center}
\vspace{0.3cm}
\begin{center}
{
\large
M.C.~David Marsh$^1$, Helen R.~Russell$^2$, Andrew C.~Fabian$^2$,\\ Brian R.~McNamara$^{3,\, 4}$,   Paul Nulsen$^{5,\, 6}$, Christopher S.~Reynolds$^{7,\, 8 }$
}
\end{center}

\vspace{0.1cm}

\begin{center}
\vskip 4pt
\textsl{
$^1$ Department of Applied Mathematics and Theoretical Physics, \\ University of Cambridge, Cambridge,  United Kingdom}\\
\textsl{$^2$ Institute of Astronomy, University of Cambridge,  Cambridge, United Kingdom} \\
\textsl{$^3$ Department of Physics and Astronomy, University of Waterloo, Waterloo,  Canada} \\
\textsl{$^4$ Perimeter Institute for Theoretical Physics, Waterloo, Canada} \\
\textsl{$^5$ Harvard-Smithsonian Center for Astrophysics, 
Cambridge,  USA \\
$^6$ ICRAR, University of Western Australia, Perth, Australia} \\
$^7$ Department of Astronomy, University of Maryland, College Park,  USA \\
$^8$ Joint Space-Science Institute (JSI), College Park, USA
\emailAdd{m.c.d.marsh@damtp.cam.ac.uk}
\emailAdd{hrr27@ast.cam.ac.uk}

\end{center} 
{\small  \noindent  \\[0.2cm]
\noindent
Axion-like particles (ALPs) and photons can quantum mechanically interconvert when propagating through magnetic fields, and ALP-photon conversion may induce oscillatory features in the spectra of astrophysical  sources. We use deep (370 ks), short frame time \textit{Chandra} observations of the bright nucleus at the centre of the radio galaxy M87 in the Virgo cluster to search for signatures of light ALPs. The absence of substantial irregularities in the X-ray power-law spectrum leads to a new upper limit on the photon-ALP coupling, $\g$: using a very conservative model of the cluster magnetic field consistent with Faraday rotation measurements from M87 and M84, 
 we find
 \be  
 \g < 2.6 \times 10^{-12}~{\rm GeV}^{-1}  \nonumber
 \ee
 at 95$\%$ confidence level for ALP masses $m_a \leq 10^{-13}$ eV. Other consistent magnetic field models lead to stronger limits of  $\g \lesssim 1.1$--$1.5 \times 10^{-12}~{\rm GeV}^{-1}$. 
  These bounds are  all stronger than the limit inferred from the absence of a gamma-ray burst from SN1987A, and  rule out a substantial fraction of the parameter space accessible to future experiments such as ALPS-II and IAXO.
 
 }

\vfill
\begin{flushleft}
\small \today
\end{flushleft}
\end{titlepage}

\newpage
\tableofcontents

\section{Introduction}
Symmetry is key  in the modern understanding of fundamental physics, and 
the universe 
may support additional symmetries that have yet to be discovered. A promising route to search for  new symmetries is to search for the remnants 
of broken symmetries.
  A spontaneously broken exact (global) symmetry results in a massless scalar particle called a Nambu-Goldstone boson; if the symmetry is only approximate, the resulting particle is a naturally light pseudo-Nambu-Goldstone boson \cite{Weinberg:1972fn}. A well-known example of a pseudo-Nambu-Goldstone boson is the QCD axion arising from the breaking of a new, postulated  $U(1)$ Peccei-Quinn symmetry \cite{Peccei:1977hh, Weinberg:1977ma, Wilczek:1977pj}. The axion may explain the absence of detectable amounts of CP violation from the strong interactions, thereby solving the the `strong CP-problem'.
  
Axion-like particles (ALPs), here denoted $a$, are a class of pseudo-Nambu-Goldstone bosons that couple to electromagnetism through the interaction,
\be
{\cal L}_{\gamma a} =  -\frac{\g}{4}a F_{\mu \nu} \tilde F^{\mu \nu} = \g\, a\, \vec{E} \cdot \vec{B} \, ,
\label{eq:Lmix}
\ee
where the ALP-photon coupling $\g$ has dimension of inverse energy. 
Much of the low-energy phenomenology of ALPs is captured by only two parameters \cite{Sikivie:1983ip, Raffelt:1987im}: $\g$ and the  ALP mass $m_a$, which can naturally be small compared to other particle physics mass scales. 

In the presence of background magnetic fields, equation \eqref{eq:Lmix} induces quantum mechanical oscillations between photons and ALPs. Generically, the stronger and more coherent the magnetic field, the larger the probability of interconversion. Very strong magnetic fields that are coherent over laboratory scales are used in many experimental searches for axions (for a recent review, see \cite{Graham:2015ouw}). Astrophysical magnetic fields that are coherent over substantially  larger scales have been used to derive strong constraints on the axion-photon coupling. 
For example, the absence of an associated gamma-ray burst from supernova 1987A (SN1987A)  has been shown to imply $\g \lesssim 5.3 \times 10^{-12}$\, GeV$^{-1}$ for light ALPs with $m_a \lesssim 4.4 \times 10^{-10}$ eV \cite{Brockway:1996yr, Grifols:1996id, Payez:2014xsa}.  
For reviews of experimental and astrophysical axion and ALP bounds, see \cite{Raffelt:1996wa, Essig:2013lka, Ringwald:2012hr}.


Galaxy clusters are extremely efficient converters of ALPs and photons with energies $\omega \gtrsim {\cal O}({\rm keV})$ \cite{Conlon:2013txa, Angus:2013sua, Conlon:2015uwa, Powell:2014mda, Cicoli:2014bfa} (see also \cite{Fairbairn:2009zi, Burrage:2009mj, Horns:2012kw, Wouters:2013hua, Wouters:2012qd, DAmico:2015snf,Schlederer:2015jwa}). The conversion probability is energy dependent and exhibits quasi-oscillatory features \cite{Wouters:2012qd, Conlon:2015uwa}.  ALP-photon conversion then leads to energy-dependent distortions of the photon spectrum of astrophysical sources, e.g.~active galactic nuclei (AGN), located within or behind a galaxy cluster. 
The absence of large spectral distortions in the spectra of localised sources 
can  be used to constrain the axion-photon coupling, \g. 
Applying this approach to \textit{Chandra} observations of the central AGN of the Hydra A cluster, reference
 \cite{Wouters:2013hua} 
 found the bound $g_{a \gamma} < 8.3 \times 10^{-12}$ GeV$^{-1}$ for $m_a < 7 \times 10^{-12}$ eV. More recently, reference \cite{Berg:2016ese} used \textit{Chandra} observations of the AGN NGC1275 in the Perseus cluster to derive the bound $g_{a \gamma} \lesssim 3.8$--$5.9 \times 10^{-12}$ GeV$^{-1}$ for $m_a \lesssim 10^{-12}$ eV.\footnote{Optimistic models of the cluster magnetic field then lead to stronger bounds. For NGC1275 in Perseus, some magnetic field models give $\g \lesssim 2 \times 10^{-12}\, {\rm GeV}^{-1}$ \cite{Berg:2016ese}.  } However, neither of these sources are ideal for deriving constraints on axion-like particles from existing observational data: Hydra A is a rather distant cluster, and 
possible modulations due to ALPs are hard to disentangle from 
statistical fluctuations and 
absorption features induced by the 
dense gas close to the source. 
For \textit{Chandra} observations of NGC1275, the nuclear spectrum is  piled up (strongly so, for a substantial fraction of the observations) with multiple photons arriving at the same detector location in the same frame integration time.  These photons are then confused with a single, more energetic photon or potentially a cosmic ray and then either excluded by the data reduction pipeline or not telemetered to the ground from the satellite.  
This makes it challenging to unambiguously associate spectral features with new, non-standard physics. Moreover, the magnetic field in both the Perseus and Hydra A clusters are not well known, which makes the rate of ALP-photon conversion uncertain \cite{Berg:2016ese}. 

In this paper, we derive constraints on \g~from the absence of large distortions of the spectrum of the the central AGN of M87 in the  Virgo cluster. This source is ideal for several reasons: at a distance of $16.1$ Mpc, Virgo is the closest galaxy cluster  to earth.  This has made precision studies of the gas structure of Virgo possible, and \textit{ROSAT}, \textit{XMM-Newton} and \textit{Chandra} have  mapped out the electron density of the cluster with  high resolution \cite{Nulsen, Urban:2011ij,2015MNRAS.451..588R}. 
The central galaxy of the Virgo cluster, M87, hosts a supermassive black hole which appears as a very bright central X-ray point source and a jet. 
 M87 was observed for a large \textit{Chandra} program with a short frame time to mitigate the impact of pile-up and provide a very deep, clean spectrum of the central AGN.
 Moreover, 
the magnetic field in Virgo has been studied using rotation measurements, indicating a strong cluster magnetic field with a central value of $B_0 \approx 35$--$40\, \mug$ \cite{Owen, Algaba, Giudetti}. As we will show, 
%
this makes ALP-photon conversion unsuppressed at X-ray energies for a significant portion of the available parameter space. In combination, these factors allow us to derive the strongest bounds to date on light axion-like particles from \textit{Chandra} observations of M87.  

Under conservative assumptions of the magnetic field structure, we find that the absence of large irregularities in the X-ray spectrum 
of M87
leads to 
a  new upper  bound on \g~for ALPs with $m_a < 3\times 10^{-12}\, {\rm eV}$. For very light ALPs with $m_a<10^{-13}\, {\rm eV}$, we find,
\be
\g \big|_{B_0=31.5{\rm \mu G}} < 1.5\times10^{-12}~{\rm GeV}^{-1}~~~{\rm (at~95\%~c.l.) } \, .
\ee
We verify that this  bound is
 insensitive to uncertainties in the cluster background subtraction. Modifications of the cluster magnetic field model (including the central magnetic field strength and its radial fall-off) can change the derived limit by up to $\sim75\%$; we consider several modifications to the magnetic field model and find that the weakest limit on $\g$ to be given by, 
 \be
 \g \big|_{B_0=35{\rm \mu G}} < 2.6\times10^{-12}~{\rm GeV}^{-1}~~~{\rm (at~95\%~c.l.) } \, .
 \ee 
  
 This paper is organised as follows:  in section \ref{sec:2} we 
discuss the observational data used in this analysis, and in section \ref{sec:3} we discuss our models for the cluster magnetic field and electron density. 
 In section \ref{sec:4} we review the physics of ALP-photon conversion, and in section \ref{sec:5} we present our new parameter constraints.
We present our conclusions  in section \ref{sec:concl}. In Appendix \ref{app:A}, we compare the Bayesian parameter constraints with those obtained using the frequentist's profile likelihood method, and we discuss the sensitivity of the constraints to the cluster gas properties and the magnetic field model.  
 
\section{M87 and the Virgo cluster}
\label{sec:2}
M87 has been observed frequently by the \textit{Chandra} X-ray
Observatory to study the nuclear and jet activity, the hot gas
atmosphere and the population of low mass X-ray binaries.  However,
the bright nuclear point source is heavily piled up in observations
using \textit{Chandra's} standard $3.1\s$ frame integration time.
Pile-up occurs when multiple photons arrive in the same detector
region within the same frame integration time and are therefore
detected as a single photon event \cite{Davis01}.  The photon energies sum to
make a detected event of higher energy, which distorts the source
spectrum, and the irregular shape of the charge cloud on the detector causes grade
migration, where events are misidentified as cosmic rays and excluded.
A shorter frame integration time of $0.4\s$ is required to reduce pile-up of the
nucleus to at most a few per cent for M87.  

The centre of M87 was observed with a short $0.4\s$ frame time for a
total of $330\ks$, split over nine observations, as a Cycle 17 Large
Project (PI Russell).  In addition to this new dataset, we also
utilized all $0.4\s$ frame time observations of M87 in the
\textit{Chandra} archive taken since 2010.  This results in an
additional sixteen observations each of $\sim5\ks$.  Intense flaring of
the jet knot HST-1, located only 0.85 arcsec from the nucleus,
resulted in strong pile-up even in $0.4\s$ frame time observations
taken between 2004 and 2010 \cite{Harris09}.  The observations used in
this analysis are detailed in Table \ref{tab:obs}.  Extraction of the
nuclear spectrum from the background cluster emission follows the
analysis of the \textit{Chandra} archive observations of M87 in
\cite{2015MNRAS.451..588R}, which is summarised below.

\begin{table*}
\begin{minipage}{\textwidth}
  \caption{Details of the \textit{Chandra} observations used for this
    analysis and the best-fit nuclear spectral model parameters.  The
    nuclear spectrum was fitted with an absorbed powerlaw model
    \textsc{phabs(zphabs(powerlaw))} over the energy range
    $0.5$--$7\keV$.  (1) \textit{Chandra} observation ID, (2) observation date, (3) exposure time, (4) intrinsic absorption, (5) photon index, (6) unabsorbed $2$--$10\keV$ flux.}
\begin{center}
\begin{tabular}{l c c c c c}
\hline
Obs. ID & Date & Exposure & Intrinsic $n_{\mathrm{H}}$ & $\Gamma$ & Flux ($2-10\keV$)\\
 & 
 & (ks) &  ($10^{22}\pcmsq$) & & 
 ($10^{-12}\ergpcmsqps$) \\
\hline
11513 & 13/04/2010 & 4.7 & $0.04\pm0.02$ & $2.33\pm0.08$ & $1.42^{+0.10}_{-0.09}$ \\
11514 & 15/04/2010 & 4.5 & $0.03\pm0.02$ & $2.14\pm0.09$ & $1.4\pm0.1$ \\
11515 & 17/04/2010 & 4.7 & $0.02\pm0.02$ & $2.20\pm0.08$ & $1.6\pm0.1$ \\
11516 & 20/04/2010 & 4.7 & $<0.02$ & $2.02^{+0.08}_{-0.07}$ & $1.6^{+0.09}_{-0.11}$ \\
11517 & 05/05/2010 & 4.7 & $0.07\pm0.02$ & $2.35\pm0.08$ & $1.5\pm0.1$ \\
11518 & 09/05/2010 & 4.0 & $0.04\pm0.02$ & $2.3\pm0.1$ & $1.15^{+0.10}_{-0.09}$ \\
11519 & 11/05/2010 & 4.7 & $0.06\pm0.02$ & $2.4\pm0.1$ & $1.01\pm0.08$ \\
11520 & 14/05/2010 & 4.6 & $0.09\pm0.02$ & $2.5\pm0.1$ & $1.00\pm0.08$ \\
13964 & 05/12/2011 & 4.5 & $0.05^{+0.03}_{-0.02}$ & $2.3\pm0.1$ & $1.16\pm0.09$ \\
13965 & 25/02/2012 & 4.6 & $0.02\pm0.02$ & $2.10^{+0.09}_{-0.08}$ & $1.4\pm0.1$ \\
14973 & 12/03/2013 & 4.4 & $0.03\pm0.02$ & $2.3\pm0.1$ & $1.16\pm0.09$ \\
14974 & 12/12/2012 & 4.6 & $0.03\pm0.03$ & $2.2\pm0.1$ & $1.15^{+0.10}_{-0.09}$ \\
16042 & 26/12/2013 & 4.6 & $<0.02$ & $2.10^{+0.08}_{-0.05}$ & $1.02^{+0.08}_{-0.07}$ \\
16043 & 02/04/2014 & 4.6 & $<0.02$ & $2.07^{+0.07}_{-0.05}$ & $1.57^{+0.10}_{-0.09}$ \\
17056 & 17/12/2014 & 4.6 & $0.04\pm0.03$ & $2.3\pm0.1$ & $1.04^{+0.09}_{-0.08}$ \\
17057 & 19/03/2015 & 4.6 & $0.03\pm0.03$ & $2.01\pm0.09$ & $1.6\pm0.1$ \\
18232 & 27/04/2016 & 18.2 & $0.08\pm0.02$ & $2.29\pm0.07$ & $0.70^{+0.04}_{-0.03}$ \\
18233 & 23/02/2016 & 37.2 & $0.03\pm0.01$ & $2.27\pm0.05$ & $0.62\pm0.02$ \\
18781 & 24/02/2016 & 39.5 & $0.04\pm0.01$ & $2.21\pm0.05$ & $0.67\pm0.02$ \\
18782 & 26/02/2016 & 34.1 & $0.04\pm0.02$ & $2.21\pm0.05$ & $0.70\pm0.03$ \\
18783 & 20/04/2016 & 36.1 & $0.05\pm0.02$ & $2.28\pm0.05$ & $0.55\pm0.02$ \\
18836 & 28/04/2016 & 38.8 & $0.08\pm0.02$ & $2.27\pm0.05$ & $0.72\pm0.02$ \\
18837 & 30/04/2016 & 13.7 & $0.04\pm0.03$ & $2.32\pm0.09$ & $0.52^{+0.04}_{-0.03}$ \\
18838 & 28/05/2016 & 56.3 & $0.04\pm0.01$ & $2.30\pm0.05$ & $0.50\pm0.02$ \\
18856 & 12/06/2016 & 24.5 & $0.05\pm0.02$ & $2.30\pm0.07$ & $0.50^{+0.03}_{-0.02}$ \\
\hline
\end{tabular}
\end{center}
\label{tab:obs}
\end{minipage}
\end{table*}

\subsection{Data reduction}
\label{sec:reduction}


All datasets were analysed with \textsc{ciao} 4.8 and \textsc{caldb}
4.7.2 supplied by the \textit{Chandra} X-ray Center
\cite{Fruscione06}.  Level 1 event files were reprocessed for the
latest gain and charge transfer inefficiency correction and then
filtered to remove cosmic rays.  Background light curves were
extracted from each observation in a region free of point sources.
These light curves were compared to identify any time periods with
high background count rates due to flares.  No major flares were found
in any of the observations.  The final cleaned exposure times are
detailed in Table \ref{tab:obs}.  The total cleaned exposure time is
$371.5\ks$.  Blank-sky backgrounds were also generated for each
dataset.  The appropriate background dataset was reprocessed similarly
to the corresponding observation, reprojected to the correct sky
position and normalized to match the observed count rate in the
$9.5-12\keV$ energy band.  Point sources were identified using a hard
energy band in each separate observation and a combined image and
excluded as necessary.  The emission from the jet was also carefully excluded.


The cluster background accounts for up to 10\% of the emission within
the region covered by the nuclear point spread function.
Reference \cite{2015MNRAS.451..588R} showed that the cluster surface brightness increases
towards the nucleus in M87, therefore a background spectrum extracted
in a surrounding annulus will likely underestimate the true
background.  Following \cite{2015MNRAS.451..588R}, we have therefore used a
ChaRT/\textsc{marx} PSF simulation to subtract the nuclear emission
and evaluate the effect of the observed increase in the background cluster surface
brightness.  The stellar population and unresolved low mass X-ray
binaries contribute at most a few per cent of this background and are
not considered further \cite{Revnivtsev08}.

\begin{figure*}
\centering
\begin{minipage}{\textwidth}
\includegraphics[width=0.45\textwidth]{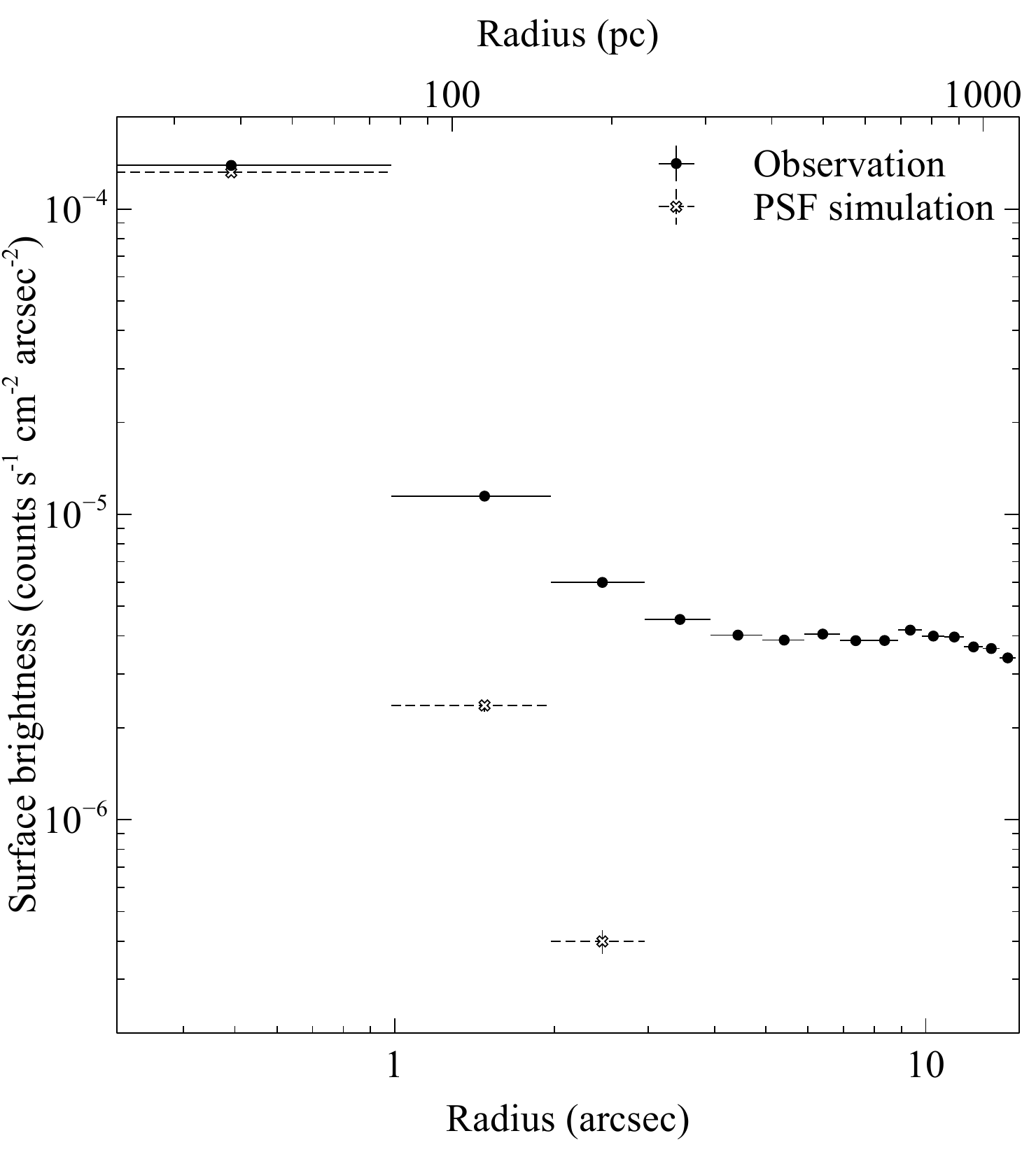}
\includegraphics[width=0.45\textwidth]{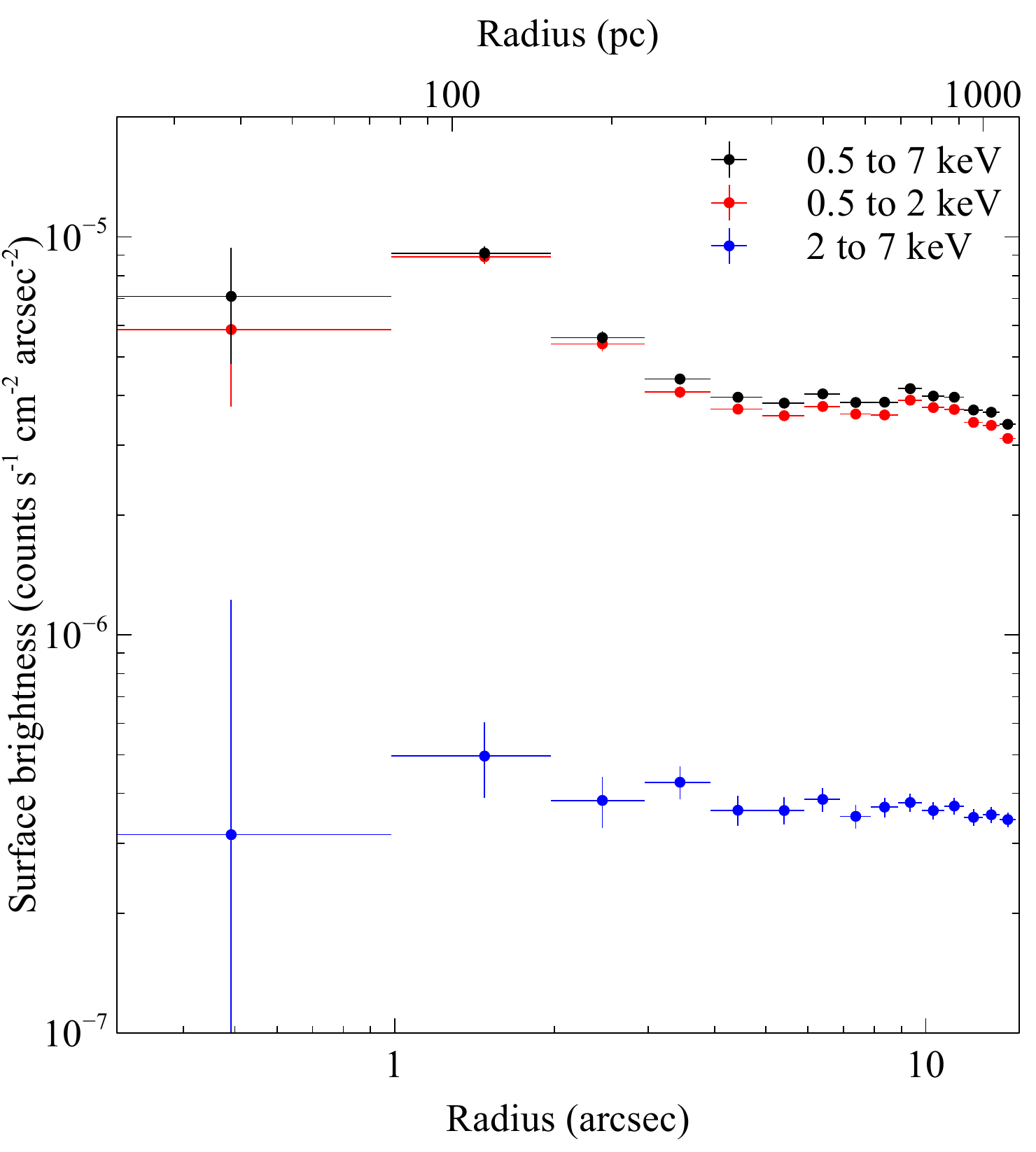}
\caption{Left: Blank-sky background subtracted surface brightness profile from obs. ID 18838 together with a ChaRT/\textsc{marx} simulation of the nuclear PSF.  Both profiles used a $0.5$--$7\keV$ energy band.  Right:  PSF-subtracted cluster surface brightness profiles in the energy bands $0.5$--$7\keV$, $0.5$--$2\keV$ and $2$--$7\keV$.}
\label{fig:chartfigs}
\end{minipage}
\end{figure*}


The nuclear flux in M87 has decreased over time, therefore the cluster
background contribution is most significant for the most recent
datasets.  We therefore analysed the cluster contribution in a recent
observation with the lowest nuclear flux (obs. ID 18838), which is
also the longest individual observation.  The spectrum of the nuclear
point source was extracted in this observation using a circular region
centred on the emission peak with a radius of 1 arcsec.  An
appropriate response and ancillary response was also generated.  The
cluster background was estimated using an annulus covering a radial
region from 2 to 4 arcsec centred on the point source.  The high
energy wings of the \textit{Chandra} point spread function (PSF)
prevent the use of a background region closer to the bright nucleus
(\cite{Jerius02}).  The spectrum was restricted to the energy range
$0.5$--$7\keV$, grouped with a minimum of 20 counts per spectral bin and
analysed in \textsc{xspec} version 12.9.0 \cite{Arnaud96}.  An
absorbed powerlaw model \textsc{phabs(zphabs(powerlaw))} was fitted to
the spectrum where the two absorption components account for the
Galactic foreground absorption and the intrinsic absorption local to
M87, respectively.  The Galactic absorption was fixed to the measured
HI column density of $1.94\times10^{20}\pcmsq$ (\cite{Kalberla05}).
The best-fit photon index was $2.30\pm0.05$, the intrinsic absorption
was $\left(4\pm1\right)\times10^{20}\pcmsq$ and the flux in the energy range
$2$--$10\keV$ was $\left(0.50\pm0.02\right)\times10^{-12}\ergpcmsqps$.  Table
\ref{tab:obs} shows the results of equivalent analyses for each of the
observations of M87 to demonstrate the decrease in nuclear flux and
the consistency of the best-fit absorption and photon index
parameters.  Observation 18838 should therefore provide a conservative representation of
the larger dataset.

From the best-fit spectral model and the source position on the
detector, the ChaRT ray-tracing program \cite{Carter03} and the
\textsc{marx} software (version 5.3.2, \cite{Davis12}) were used to
simulate an observation of the nuclear PSF.  ChaRT produces ray-traces
through the \textit{Chandra} mirror model and \textsc{marx} projects
these ray-traces onto the ACIS-S detector.  This combination provides
an accurate simulation of the on-axis \textit{Chandra} PSF to at least
10 arcsec \cite{Jerius02}.  Figure \ref{fig:chartfigs} (left) shows the
observed surface brightness profile and the corresponding
ChaRT/\textsc{marx} simulation of the nuclear contribution.  The
simulated nuclear surface brightness profile was subtracted from the
observed profile to reveal the background cluster contribution.
Figure \ref{fig:chartfigs} (right) shows that the cluster surface
brightness measured in a 2--4 arcsec region will underestimate the
cluster background in the nuclear region within 1 arcsec.  However,
this increase in surface brightness is limited to the soft energy band
(0.5--2$\keV$), which traces the cooler, denser X-ray gas at the
cluster centre.  The harder energy band at 2--7$\keV$ primarily traces
hotter, projected gas and therefore has a flat surface brightness
profile.  The ALP models considered here are primarily constrained in
the 2--7$\keV$ energy range.  Therefore, the cluster background
determined from 2 to 4 arcsec is sufficient.  We test the effect of
the uncertainty in the cluster background in section \ref{sec:bkgrnd}.

\subsection{Nuclear X-ray spectrum}

\begin{figure*}
\centering
\begin{minipage}{\textwidth}
\includegraphics[width=0.45\textwidth]{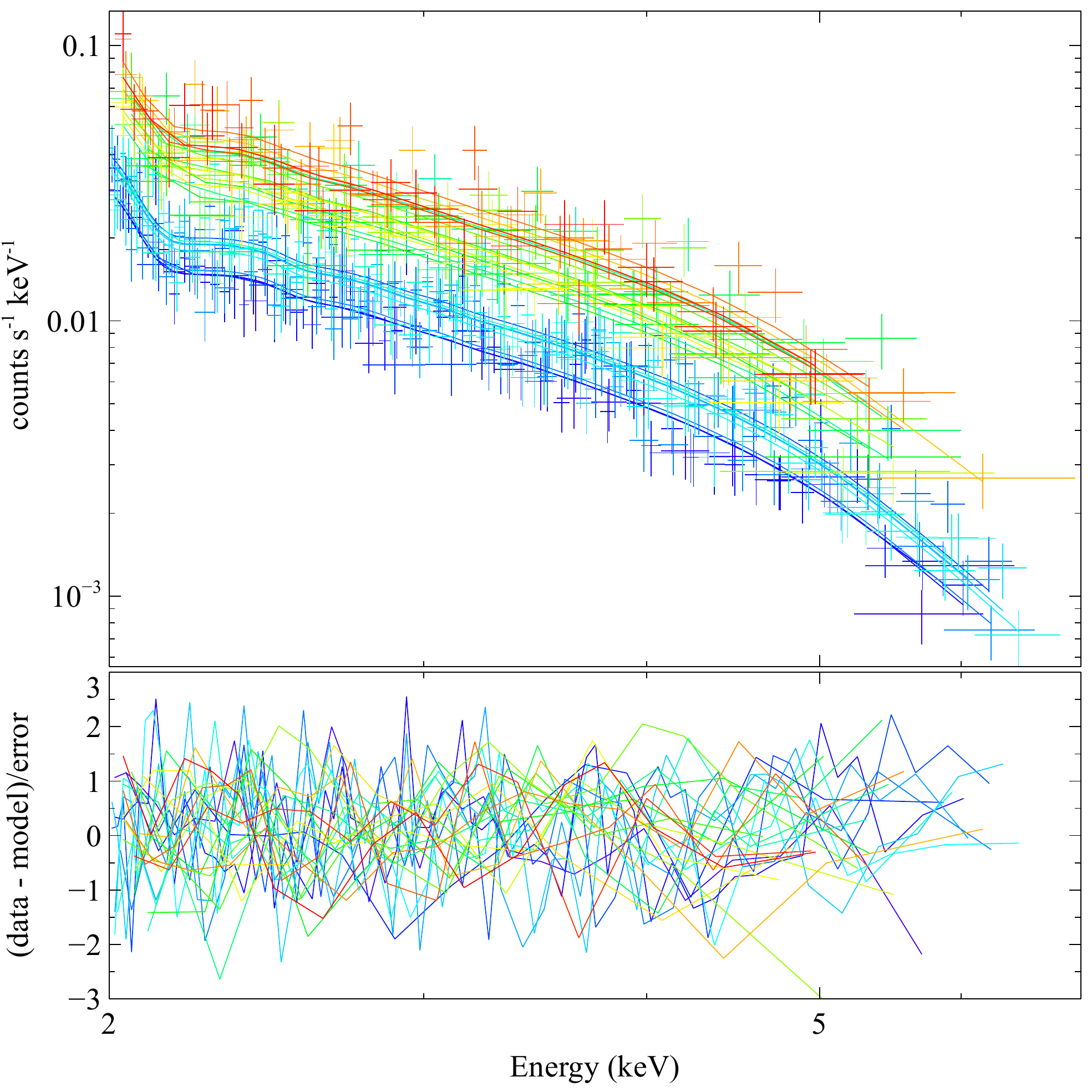}
\includegraphics[width=0.45\textwidth]{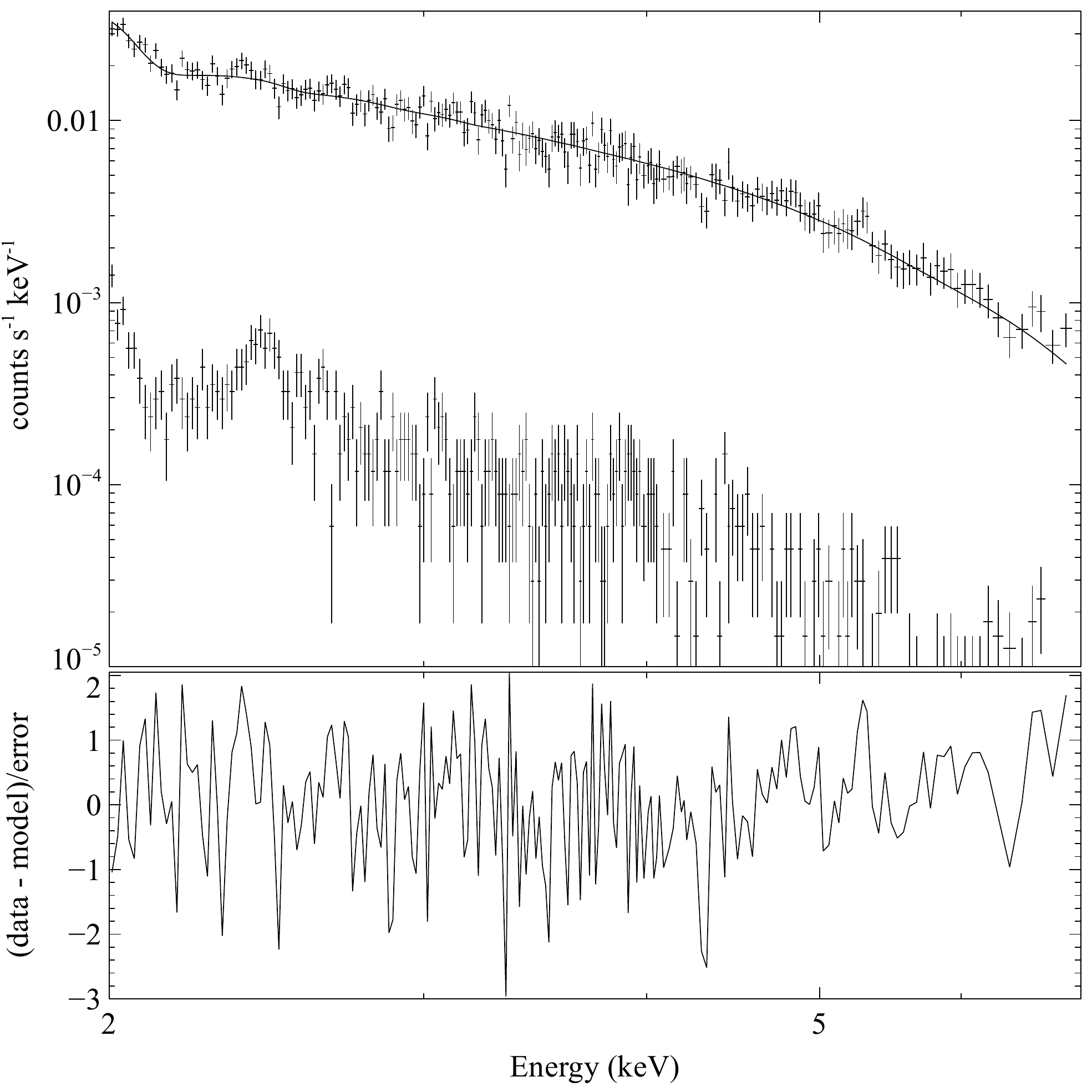}
\caption{Left: Individual nuclear spectrum extracted from each of the 25 observations.  Right: Summed spectrum of all observations taken in 2016 to demonstrate the data quality.  The total exposure time is $298.4\ks$.  The background spectrum is also shown.}
\label{fig:spectra}
\end{minipage}
\end{figure*}

For the ALP model analysis, the extracted nuclear spectrum and corresponding background spectrum for each observation  were fitted together in \textsc{xspec} with an absorbed
powerlaw model \textsc{phabs(zphabs(powerlaw))} over the energy range 2--7$\keV$.  The absorption
components were fixed to the Galactic value and the best-fit intrinsic
absorption value of $3.9\times10^{20}\pcmsq$ \cite{2015MNRAS.451..588R}.  The
$\chi^2$ statistic was used to determine
the best-fit model.  The photon index parameter was tied between the
observations whilst the normalization was left free to account for the
flux variability.  The absorption and photon index do not appear to
significantly vary over the six year time span of these observations (Table \ref{tab:obs}).
Fig. \ref{fig:spectra} (left) shows the individual nuclear spectra
from each observation.  The best-fit photon index is $2.36\pm0.03$,
where $\chi^2=597.4$ for 614 degrees of freedom.  Fig. \ref{fig:spectra} shows the nuclear spectrum summed over the
observations taken in 2016 to more clearly demonstrate the data
quality and the spectral fit.  \textit{Chandra's}
effective area has changed significantly over the mission lifetime,
therefore we can only sum observations taken at similar epochs.  This summed spectrum and the
corresponding spectral fit were therefore not used in our subsequent analysis.
In section \ref{sec:5}, we add the ALP-photon conversion model to the
absorbed power-law model to place constraints on the ALP parameters
$m_a$ and $\g$.

\section{Modelling the Virgo cluster}
\label{sec:3}
Being nearby, bright and astrophysically rich, the Virgo cluster has been studied in great detail at radio, optical and X-ray wavelengths. As we discuss in this section, this makes it possible to construct observationally well-constrained models of the cluster magnetic field and gas density.  In section \ref{sec:nevirgo}, we briefly review existing \textit{ROSAT}, \textit{XMM-Newton} and \textit{Chandra} determinations of the electron density of the intracluster gas, and in sections \ref{sec:Bvirgo} and \ref{sec:Bmodel}
 we discuss how multiple radio observations of the radio galaxies M87 and M84 can be used to constrain the galaxy cluster magnetic field.



\subsection{The gas density distribution}
\label{sec:nevirgo}
The Virgo cluster is a `cool core' cluster for which the cooling time of the X-ray emitting gas at the cluster centre is considerably smaller than the Hubble time, leading to a highly peaked surface brightness profile and the existence of a multi-temperature plasma at the centre of the cluster \cite{Stewart1984, Fabian1984}. The power output of the AGN at the centre of M87  sources a complex astrophysical environment with radio emitting lobes extending to distances of $\sim 40\, {\rm kpc}$ from the jet (eg. \cite{Young02,FormanM8705,FormanM8707,deGasperin12}). 

The gas density distribution has been studied by several X-ray satellites: reference \cite{Nulsen} used \textit{ROSAT} observations to determine the electron density within the central $\sim 300\,{\rm kpc}$ region of M87; more recently, reference \cite{Urban:2011ij} collated 13 \textit{XMM-Newton} pointings covering, on large scales,  the Virgo cluster from its centre and northwards to radii beyond the cluster virial radius, $R_{\rm vir} = 1.08\, {\rm Mpc}$; moreover, reference \cite{2015MNRAS.451..588R} used short frame time \textit{Chandra} observations of M87 to determine the detailed gas density within the complicated, central 2 kpc region of M87 and resolve the  gas density within the Bondi radius ($r\sim 0.2{\, \rm kpc}$) of the central black hole.  

\begin{figure}
\centering
\includegraphics[width=.65\textwidth]{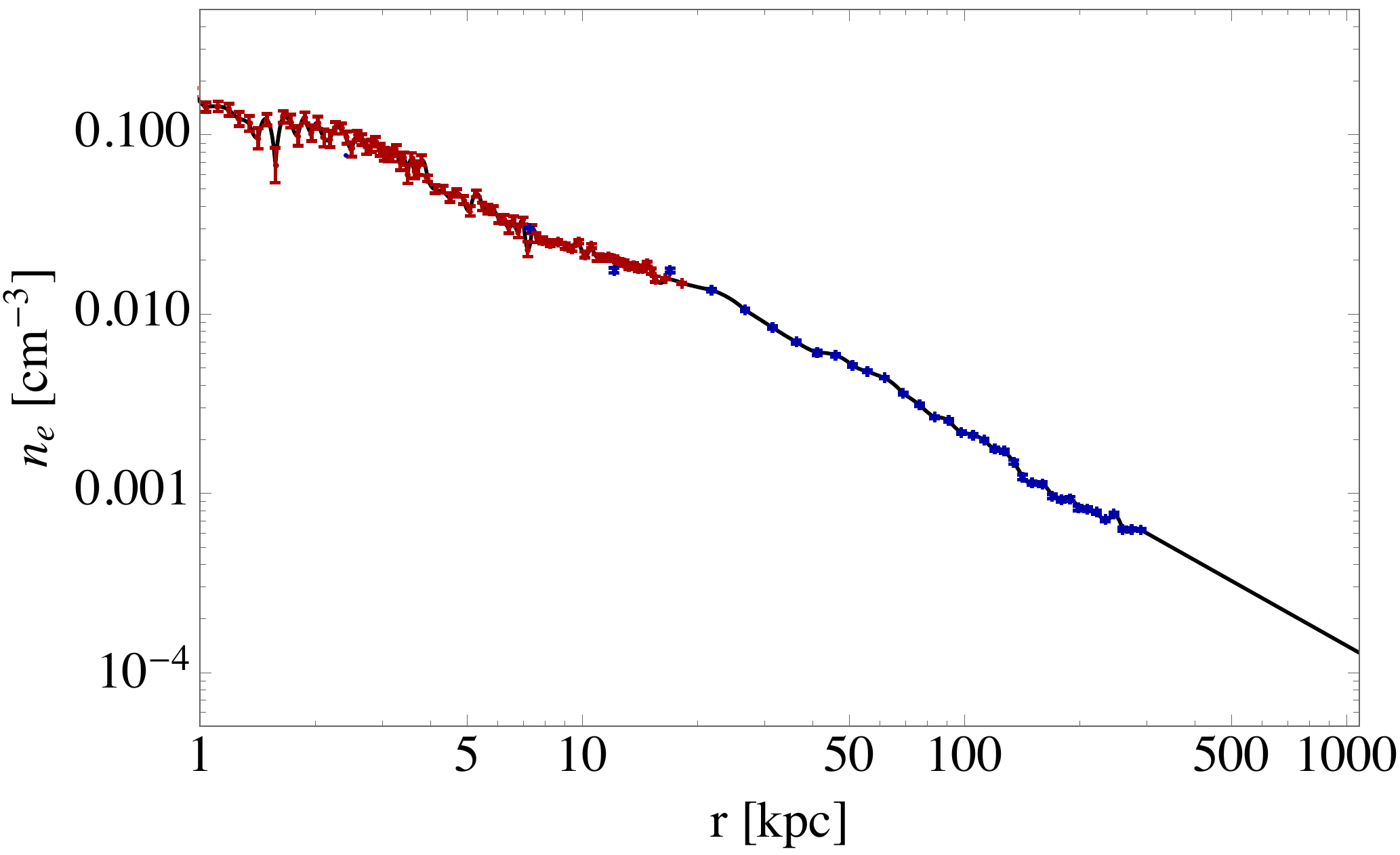}
\caption{The Virgo deprojected electron density as observed by \textit{Chandra} (red) and \textit{ROSAT} (blue) \cite{Nulsen}. Our model electron density (black line) is consistent with the large-radius scaling inferred from multiple \textit{XMM-Newton} observations with $r \leq R_{\rm vir}$ \cite{Urban:2011ij}. }
\label{fig:ne}
\end{figure}

To model the Virgo cluster electron density, we  use an interpolating spline model fitted to 
\textit{Chandra} data obtained from \cite{2015MNRAS.451..588R} for $r< 19\, {\rm kpc}$, 
  \textit{ROSAT} data from \cite{Nulsen} in the $19{\, \rm kpc}< r <  298\, {\rm kpc}$ region, and  at large radii, $298 \, {\rm kpc} < r< 1080 \, {\rm kpc}$,  we assume the scaling   
 $n_e(r) \sim r^{-1.2}$ found by \cite{Urban:2011ij} to be accurate for for $r\gtrsim100\, {\rm kpc}$. 
 Figure \ref{fig:ne} shows the combined  \textit{Chandra} and  \textit{ROSAT} data together with the model of the electron density used in this paper.

X-ray images of the Virgo cluster reveal intrinsic asphericity on large scales, due to mergers with infalling small galaxy groups, and on small scales in M87 due to the AGN activity \cite{Bohringer1994, FormanM8705}.  However, for plausible limits on the axis ratio ($\pm40\%$), the effects of large-scale triaxiality have been shown to be small, at typically less than a few per cent (see e.g.~\cite{Piffaretti, Churazov2008}).  On scales of a few kpc, radio jets and lobes powered by the central AGN displace the hot X-ray atmosphere, which produce larger uncertainties in the density profile.  The model density profile interpolates over the affected regions (e.g.~at a radius of $1.5\, {\rm kpc}$ in Figure 3). However, as discussed in section 4.2, the ALP-photon conversion probability is suppressed in the high-density cluster core, and asphericity in this region does not affect our bounds. 
In the bulk of the cluster, the uncertainty stemming from cluster asphericity is less important than
the uncertainty in the cluster magnetic field, as we will now discuss.  

\subsection{The Virgo magnetic field}
\label{sec:Bvirgo}
 A magnetised plasma, such as the Virgo intracluster medium, is birefringent and induces a wavelength-dependent rotation of the plane of polarisation of linearly polarised photons traversing it. This  `Faraday rotation' provides one of the main methods for inferring the magnetic field strength of galaxy clusters. The magnitude of the Faraday rotation is linearly proportional to the `rotation measure', RM, which is given by the  line-of-sight integral, 
\be
{\rm RM} = \frac{e^3}{2\pi m_e^2} \int_{\rm l.o.s.} n_e(l) \vec{B}(l)\cdot d\vec{l} =  \frac{e^3}{2\pi m_e^2} \int_{\rm l.o.s.} n_e(l) B_{\parallel}(l)\, dl \, .
\label{eq:RM}
\ee
With the electron density
determined by X-ray observations,  the statistical properties of the  rotation measures from a source of
polarised emission can be used to constrain the magnetic field in the `Faraday screen' between the source and the observer.

Clearly, regions with a large electron density contribute comparatively more to the RM than low density regions. Moreover, for a statistically isotropic magnetic field, $B_{\parallel}$ is uniformly distributed around zero and the average RM over several line-of-sights is zero.\footnote{Conversely, a non-vanishing mean values of the RM over several sightlines indicate large-scale order in the magnetic field.}  The standard deviation $\sigma({\rm RM})$ of the rotation measures from an extended source indicates the 
 typical magnitude of the RM.
%
%
We here use observations of RMs from the radio galaxies M87 and M84 to constrain the Virgo cluster magnetic field.

Observations of M87 have revealed very large rotation measures. Using four-frequency observations with  the VLA, reference \cite{Owen} found RMs of  
${\cal O}(1000$--$2000)\, {\rm rad}\, {\rm m}^{-2}$ in the M87 lobes.  Even larger RMs, ${\cal O}(4000$--$8500)\, {\rm rad}\, {\rm m}^{-2}$, were found in a narrow filament, but the prominent jet region was found to give comparatively smaller RMs, ${\cal O}(300)\, {\rm rad}\, {\rm m}^{-2}$. The large rotation measures were interpreted as arising from a Faraday screen in front of the synchrotron radiating, radio-emitting plasma, with most of the Faraday rotation occurring within a few kpc from the source, where the magnetic field strength was estimated to $B_0 \approx 40\, \mug$. In this interpretation, the lower rotation measures from the jet can be explained by the jet extending through and beyond the main Faraday screen.\footnote{Rotation measures from the  M87 jet has been studied by several groups.  Using VLBA observations at seven frequencies, reference \cite{Zavala} found rotation measures of ${\cal O}(9000)\, {\rm rad}\, {\rm m}^{-2}$ on milli-arc second scales. The extremely large RMs were interpreted as arising from very dense gas clouds near the central AGN. The M87 jet has further been studied on kiloparsec scales in reference \cite{Algaba} using four-frequency VLA data, where it was argued that the RMs in the jet is of a different origin than the RMs of the lobes. }  

In reference \cite{Giudetti}, VLA data was used to infer the rotation measures of M87 with 0.4 arcsecond resolution, finding results largely consistent with reference \cite{Owen}: the mean RM was found to range from around $-1000$ to $6000$ rad$\, {\rm m}^{-2}$, with typical values of  ${\cal O}(1000)\, {\rm rad}\, {\rm m}^{-2}$. Very large rotation measures ${\cal O}(3500$--$6000)\, {\rm rad}\, {\rm m}^{-2}$ were only found in a small region, and the jet was again found to be characterised by lower RMs ${\cal O}(100)\, {\rm rad}\, {\rm m}^{-2}$. By approximating the RMs as a Gaussian, isotropic variable, 
the RM power spectrum was found to be flatter than anticipated by Kolmogorov turbulence. The magnetic field autocorrelation length was estimated to 0.2 kpc and the central magnetic field strength to ${\cal O}(35\, \mug)$. 

The large rotation measures observed in M87 should be contrasted with those observed in the radio galaxy M84, located around $400$ kpc from the centre of Virgo. In reference \cite{Laing}, the magnitude of these were found to be 25--35  ${\rm rad}\, {\rm m}^{-2}$. Also in this case the origin of the rotation measures was interpreted to arise from the magnetised plasma within a few kpc from the source, as opposed to the bulk of Virgo cluster. We will find in section \ref{sec:Bmodel} that these observations provide highly complementary constraints on the cluster magnetic field to those inferred from rotation measures of M87. 

We note in closing that cluster magnetic fields may in general have structure on scales $\ll {\, \rm kpc}$ which can be hard to infer from rotation measures. For example, kinetic-numerical simulations indicate that mirror and firehose instabilities driven by shear in  the turbulent
cluster plasma can generate  field fluctuations of order $\delta B
/ B_0 \simeq 1$ on scales comparable to the ion Larmor
radius \cite{Kunz}, which is microscopic compared to the coherence scales inferred from rotation measures. 
As we explain in section \ref{sec:spectdist}, such  small-scale structure is likely to have a negligible impact on the ALP-photon conversion rate from the cluster, and we do not attempt to capture these effects in our magnetic field model, which we now describe.  





\subsection{Magnetic field model}
\label{sec:Bmodel}
The complicated, tangled magnetic field of the Virgo cluster cannot be 
directly determined
 from astronomical observations, however, it is possible to construct models of the cluster magnetic field that are consistent with the observed Faraday rotation measures.

We model the cluster magnetic field along a given line-of-sight (that we here take to extend radially from the centre of the cluster) to consist of multiple domains of constant magnetic field,
\be
\vec{B}(r) = \sum_{i=0}^N (\vec{B})_i(r) \, ,
\ee
where $(\vec{B})_i$ is only non-vanishing for $R_i < r< R_{i+1}$, and is constant over this domain. In each domain, the magnetic field is described by a random, isotropically distributed unit vector, $(\vec{b})_i$, multiplying a radial scaling function,
\be
(\vec{B})_i = B_0 \left(\frac{n_e(r)}{n_e(0)} \right)^{\alpha} (\vec{b})_i \, ,
\ee
where $B_0$ denotes the central magnetic field strength and $\alpha$ determines the fall-off of the magnetic field relative to the gas density. 

Clearly, this simple model can only be expected to provide an approximation to the multi-scale structure and radial fall-off of the actual cluster magnetic field. Furthermore,  Faraday rotation measures are only sensitive to the radial component of the magnetic field along the line-of-sight, cf.~equation \eqref{eq:RM}, while ALP-photon conversion depends on the perpendicular component, as we will discuss in section \ref{sec:4}. A substantial non-isotropic  component of the 
actual cluster magnetic
at large radii  may affect the rate of ALP-photon conversion.

\begin{figure}
\centering
\includegraphics[width=.5\textwidth]{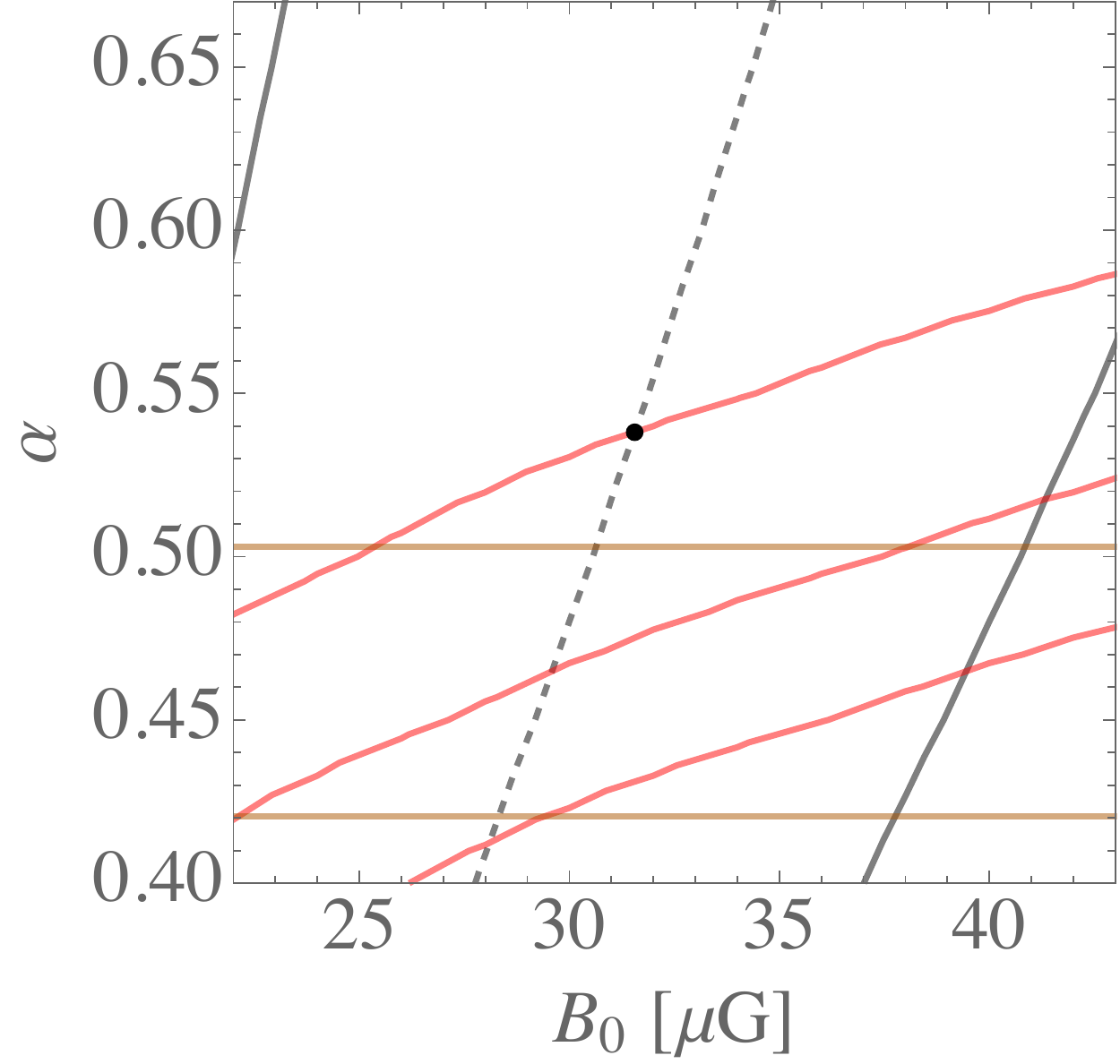}
\caption{Parameter constraints from Faraday rotation measures. From left to right, grey curves indicate $\sigma({\rm RM}_{\rm M87}) =  1000,1500\, {\rm (dashed)}, 2000\, {\rm rad}\, {\rm m}^{-2}$. From top down, red curves indicate $\sigma({\rm RM}^{\rm Virgo}_{\rm M84}) =5, 7.5, 10\, {\rm rad}\, {\rm m}^{-2}$ and orange horizontal lines indicate near-source contributions to the M87 RM of  85\% and 80\%. The black dot indicates our baseline parameters, $B_0 = 31.6\,\mug, \alpha =0.54$.}
\label{fig:Bparams}
\end{figure}

The coherence lengths, $L_i = R_{i} - R_{i-1}$, may in general depend on the distance from M87. Close to the source, 
reference \cite{Giudetti} suggests a magnetic field with an auto-correlation length of $0.2$ kpc. We here model the region within 2 kpc from the central point source as consisting of 10 domains of equal length, i.e.~$L_i = 0.2$ kpc for $1\leq i \leq 10$. Outside this region, in the bulk of the Virgo cluster, we take the coherence lengths $L_i$ to be independent and randomly distributed with a power-law probability distribution for all $i>0$,
\be
p(L_i) \sim 
\left\{
\begin{array}{l c l}
 L_i^{-\gamma} 
& ~~&{\rm for }~ L_i>L_{\rm min} 
\, , \\
0 && {\rm otherwise} \, ,
\end{array}
\right.
\ee
We extend this model to the virial radius: $R_N \approx R_{\rm virial}= 1.08$ Mpc. The  parameters of this model are then $(B_0, \alpha, \gamma, L_{\rm min})$. Motivated by more elaborate modelling of the magnetic field in cool core galaxy clusters  \cite{Vacca}, we take 
$\gamma = 2.5$ and $L_{\rm min} = 1$ kpc in our benchmark magnetic field model, and we constrain $B_0$ and $\alpha$ by the observed rotation measures  from M87 and M84.  

For the centrally located radio galaxy M87, we expect ${\rm RM}_{\rm M87} \approx {\cal O}(1000$--$2000)$. 
To determine the dependence of the rotation measures  on the magnetic field model parameters, we simulate a large number of magnetic field realisations with a given set of parameters, $(B_0, \alpha)$,  and compute the standard deviation of the rotation measures from M87:  $\sigma({\rm RM}_{\rm M87})$.
Contours of constant $\sigma({\rm RM}_{\rm M87})$ in the relevant range are indicated by grey, almost vertical lines in Figure \ref{fig:Bparams}.
According to the interpretation of reference \cite{Owen}, the Faraday rotation measures from M87 chiefly arises from the high gas density region 
with $r \lesssim r_{*} = 20$ kpc, rather than the bulk of the Virgo cluster. This constrains the radial fall-off parameter, $\alpha$. To estimate the typical near-source contribution to the rotation measure, we simulate many magnetic fields realisations for a given choice of $B_0$ and $\alpha$ and compute,
\be
{\mathfrak f} = {\rm median} \left| \frac{{\rm RM(R_{vir})} - {\rm RM(r_*)} }{{\rm RM(r_*)}} \right| \, ,
\ee
where we used the notation $
{\rm RM}(r) = \frac{e^3}{2\pi m_e^2} \int_0^r n_e(l) B_{\parallel}(l)\, dl$. Contours with ${\mathfrak f} = 0.8$ and $0.85$ are given by the orange, horizontal lines in Figure \ref{fig:Bparams}. 

The magnetic field  model is also constrained by the comparatively small rotation measures observed from M84. Similarly to the case of M87, the M84 rotation measures are expected to  chiefly arise from the magnetised plasma within $\sim 10$ kpc from the source. More elaborate models of the local magnetic field and the gas density may be used to describe this region. Here however, we expect our model to capture the (subdominant) \emph{galaxy cluster} contribution to the rotation measures from M84, ${\rm RM^{Virgo}_{M84}}$, but not the (dominant) local contribution. We may then constrain a combination of $B_0$ and $\alpha$ by placing an upper bound on  $\sigma({\rm RM^{Virgo}_{M84}})$, as illustrated by the red contours in Figure \ref{fig:Bparams}. 

These constraints from rotation measurements from M87 and M84 do not uniquely determine the parameters $B_0$ and $\alpha$. However, we note that if the energy density in the magnetic field  scales with the energy density in the plasma, then $B^2 \sim n_e$ and  $\alpha =1/2$. 
Our baseline magnetic field model used in this paper takes  $B_0 = 31.6\, \mug$, and  $\alpha = 0.54$, consistent with the constraints from rotation measures and conservatively with respect to previous estimates of the central magnetic field strength. 
%
%
We  discuss the sensitivity of ALP-photon conversion to this choice of parameters in Appendix \ref{app:A}.

\section{ALP-photon conversion}
\label{sec:4}
In this section, we describe the physics of ALP-photon conversion and our method to compute the photon survival probability for X-ray photons from M87. 

\subsection{ALP-photon conversion}
 In addition to the ALP-photon interaction \eqref{eq:Lmix}, the ALP dependent contribution to the Lagrangian is given by,
\be
{\cal L}_a = -\frac{1}{2} \partial_{\mu}a\,\partial^{\mu} a - \frac{1}{2} m^2_a\, a^2 \, .
\ee
To study ALP-photon conversion in galaxy clusters, it is convenient to linearise the equations of motion in a background with an external magnetic field, $\vec{B}({\bf x})$, and a plasma with electron density $n_e({\bf x})$. Upon neglecting the subleading effect of Faraday rotation, the equation of motion for a mode of energy $\omega$ propagating in the $z$-direction is  given by \cite{Raffelt:1987im},  
\be
\label{eq:EqofMotion}
\Big(\left(\omega - i\partial_z\right) \mathbb{I} +M(z) \Big)\left(\begin{array}{c}
								\Ket{\gamma_x} \\
								\Ket{\gamma_y} \\
								\axion
							      \end{array}\right)= 0 \, ,
\ee
where
\be
M (z) = \left(\begin{array}{c c c}
							\Delta_{\gamma}(z) & 0 & \Delta_{\gamma a x}(z) \\
							0 & \Delta_{\gamma}(z) & \Delta_{\gamma a y}(z) \\
							\Delta_{\gamma a x}(z) & \Delta_{\gamma a y}(z) & \Delta_{a}(z)
		   				\end{array}\right) \, .
						\label{eq:M}
\ee
The quantum mechanical ALP-photon oscillations are induced by the off-diagonal matrix elements,
\be
 \Delta_{\gamma a i}  =  \frac{\g B_i({\bf x})}{2} \, .
 \label{eq:Deltai}
 \ee
  Here  $\Delta_{\gamma}  = -\omega_{pl}^2/2\omega$ is determined by the plasma frequency $\omega^2_{pl}({\bf x})  = 4\pi e^2 n_e({\bf x}) /m_e$, and $\Delta_a  = -m_a^2/\omega$. 
The general solution to equation (\ref{eq:EqofMotion}) is given by the path-ordered transfer matrix,
\be
\label{eq:homsoln}
\left(\begin{array}{c}
	\Ket{\gamma_x} \\
								\Ket{\gamma_y} \\
	\axion \\
\end{array}\right)(R) =  {\cal P}_z\left[ \exp \left(-i \omega R \mathbb{I} \ -i \int_0^R M(z) \dd z\right) \right] \left(\begin{array}{c}
				\Ket{\gamma_x} \\
								\Ket{\gamma_y} \\
												\axion
		      \end{array}\right)_0 \, .
\ee
When the matrix $M(z)$ can be approximated as piece-wise constant over the ranges $R_i \leq z< R_{i+1}$ for $i = 0, \ldots, N$ (again denoting the coherence length by $L_i = R_i - R_{i-1}$), the full transfer matrix is simply given by,
\be
 {\cal P}_z\left[ \exp \left(-i \omega R \mathbb{I} \ -i \int_0^R M(z) \dd z\right) \right]
 =
 e^{-i \omega R \mathbb{I}  } \,  {\cal P}_z\left(  \prod_{i=1}^N e^{ -i L_i M(R_i) } \right) \, .
 \label{eq:transfer}
 \ee
The probability that a photon that initially is linearly polarised along the $x$-direction is converted to an ALP after propagating from $z=0$ to $z=R_N = R$ is then given by,
\be
P_{\gamma_x \to a} = \left| \left(0,0,1\right)\,  {\cal P}_z\left(  \prod_{i=1}^N e^{ -i L_i M(R_i) } \right) \left(\begin{array}{c} 1\\ 0 \\ 0 \end{array} \right) \right|^2 \, .
\ee
The conversion probability for an unpolarised beam of photons is then given by $P_{\gamma \to a} = \frac{1}{2} \left(P_{\gamma_x \to a}+ P_{\gamma_y \to a}\right)$, and the photon survival probability is given by $P_{\gamma \to \gamma} = 1- P_{\gamma\to a}$.  Since $P_{\gamma_i \to a} = P_{a \to \gamma_i }$ and $P_{a\to a} + P_{a\to \gamma_x} + P_{a\to \gamma_y} =1$, clearly    $P_{ \gamma_x \to a} + P_{\gamma_y \to a} \leq 1$ and $P_{\gamma \to \gamma } \geq 1/2$.

We close this subsection by discussing 
%
the possible impact of so-called `resonant' conversion. For ALPs more massive than the effective photon mass $\omega_{\rm pl}$, photon-to-ALP conversion tends to be suppressed by $P_{\gamma \to a} \sim \left(\omega_{\rm pl}/m_a\right)^4$. If somewhere in the cluster, $m_a\approx \omega_{\rm pl}$, the conversion probability can become much larger. To see this, we may consider $P_{\gamma \to a}$ evaluated over a single domain of size $L$ and with perpendicular magnetic field $B_\perp$:
\be
P_{\gamma \to a} = \frac{1}{4} \frac{\Theta^2}{1 + \Theta^2} \sin^2\left(
\Delta \sqrt{1+ \Theta^2}
\right) \, ,
\ee
where $\Theta = 2 B_\perp \omega \g/m^2_{\rm eff}$, $\Delta = m^2_{\rm eff} L/(4\omega)$ and $m^2_{\rm eff} = m^2_a - \omega_{\rm pl}^2$. 
If $m^2_{\rm eff} \ll \omega_{\rm pl}^2$ so that $\Theta \gg 1$, $\Delta \ll1$ with $\Theta \Delta <1$, the conversion probability can be `resonantly enhanced' to, 
\be
P_{\gamma\to a} \approx \frac{1}{4} \Theta^2 \Delta^2 = 1.2\times10^{-4} \left(\frac{B_\perp}{10\, {\rm \mu G}} \frac{L}{1\, {\rm kpc}} \frac{\g}{10^{-12}\, {\rm GeV}^{-1}} \right)^2 \, .
\label{eq:resonant}
\ee
We note that in the cluster, both $\omega_{\rm pl}$ and $B_{\perp}$ are not expected to be constant and regions in which $m^2_a \approx \omega_{\rm pl}^2$ are not very large. 
 
 In our numerical evaluation of the transfer matrix, we assume $\omega_{\rm pl}(R_i)$ and $M(R_i)$ to be piece-wise constant over a set of domains. Only a limited set ALP masses will then satisfy $m^2_a \approx \omega_{\rm pl}^2(R_i)$ for all $i$, so that our numerical implementation neglects the resonant enhancement of the amplitude of intermediate masses. However, the error associated with this effect  is negligibly small, cf.~equation \eqref{eq:resonant}, which justifies our approach.




\subsection{Spectral distortions of M87 from ALPs} 
\label{sec:spectdist}

\begin{figure*}
\centering
\begin{minipage}{\textwidth}
\includegraphics[width=0.45\textwidth]{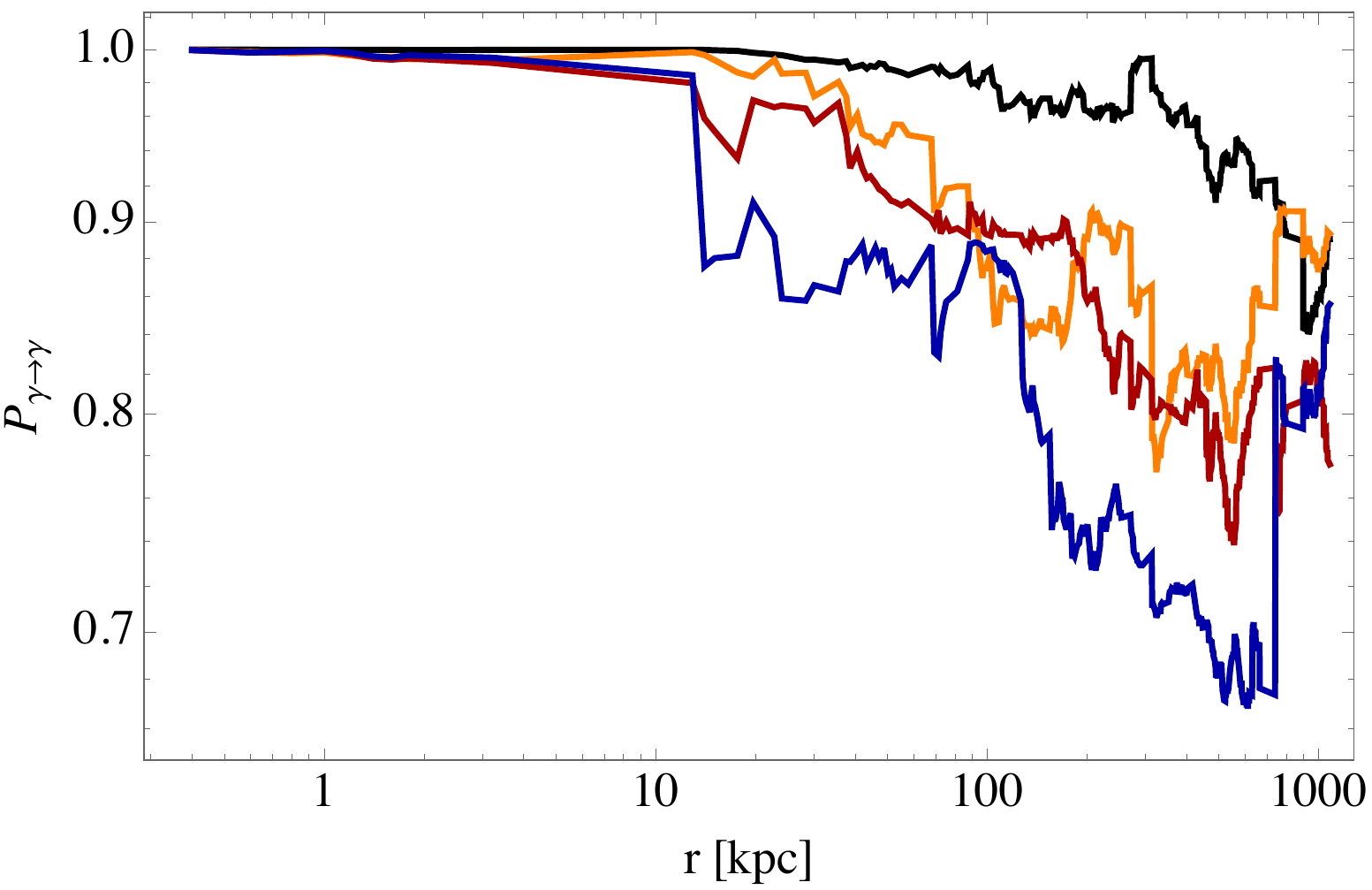}
\includegraphics[width=0.45\textwidth]{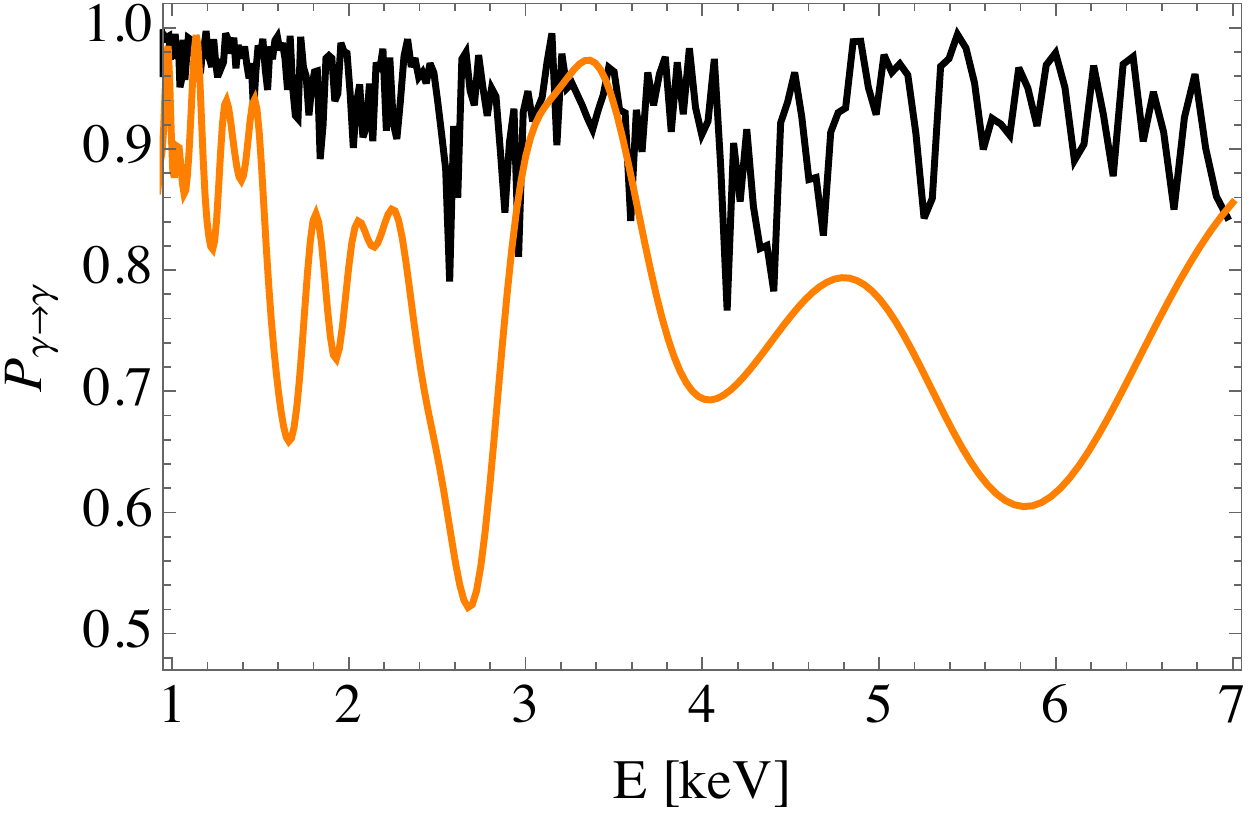}
\caption{Left: Example of radial dependence of photon survival probabilities for 
$E =1\, {\rm keV}$ (black), $3\, {\rm keV}$ (orange), $5\, {\rm keV}$ (red) and $7\, {\rm keV}$ (blue) in a
 randomly generated magnetic field with benchmark parameters, and  ALP parameters $m_a=0,~ \g = 5\times10^{-12}$\, GeV$^{-1}$.   Right: Example of energy dependence of survival probability for the same ALP parameters (orange), and for the same $\g$ but with $m_a =3.3\times10^{-12}\, {\rm eV}$ (black).}
 \label{fig:Pexpl}
\end{minipage}
\end{figure*}

Photons emitted from the bright, central source of M87 may convert into ALPs when propagating through the galaxy cluster magnetic field. In this section  we briefly summarise the phenomenology of ALP-photon conversion in the Virgo cluster.

Substantial inter-conversion of ALPs and photons occur when the off-diagonal components of the matrix \eqref{eq:M} are comparable to the diagonal terms.  For massless ALPs, $\Delta_a = 0$,  and the non-vanishing diagonal terms depend on the electron density of the environment and the energy of the particle, $\Delta_{\gamma} \sim n_e/\omega$. Hence,  large electron densities and low energies suppress the conversion probability.
Consequently, in the  cluster environment, both  the radius at which the conversion becomes important and the overall conversion probability are  energy dependent. 

Figure \ref{fig:Pexpl} illustrates how, for a  randomly generated magnetic field with  benchmark parameters, the photon-to-ALP conversion probability is highly stochastic and exhibits quasi-oscillatory features.  At small radii the conversion probability is suppressed over the entire X-ray range due to the large electron density close to the centre of the cluster. At radii $\gtrsim {\cal O}(10$--$100)\, {\rm kpc}$, ALP-photon conversion  becomes important. Higher energy photons (for which the diagonal terms of \eqref{eq:M} are more heavily suppressed) achieve large conversion probabilities at smaller radii than less energetic photons. 
%
A non-vanishing ALP mass suppresses the amplitude of the conversion and increases the frequency of the oscillations. 

It is now clear why neglecting the small-scale structure of the magnetic field is well-motivated: to leading order, the ALP-photon amplitude from the $i$:th domain scales like $\sim \g B_i L_i$, and the conversion probability scales like $\sim (\g B_i L_i)^2$. 
Domains with $L_i \ll {\rm kpc}$ provide suppressed contributions to the amplitude, and as such domains are expected to comprise a very small fraction of the cluster volume,
their effects on the conversion probability can be neglected.\footnote{Similarly, small-scale magnetic fields have a  very small effect on cluster rotation measures, which like the domain amplitude, are linear in the coherence length. } 

Finally, we note that the bright point-like source at the centre of M87 in reality has a finite extent, and photons from different parts of the source  travel along slightly different sight-lines to the detector. In general, well-separated sight-lines probe different magnetic fields, and the total cluster conversion probability becomes averaged, which suppresses the oscillatory features.  This averaging effect  becomes significant for sight-lines separated by ${\cal O}({\rm few})\, {\rm kpc}$ \cite{Conlon:2015uwa}. As the point source at the centre of M87 is not much larger than ${\cal O}(10\, {\rm pc})$ \cite{1999Natur.401..891J, 2004AJ....127..119L}, this effect is completely negligible in our case, and the single sight-line approximation is well justified.

\section{Parameter constraints}
\label{sec:5}
We are  now ready to compute the  constraints on the ALP parameters  \g~and $m_a$ from the absence of large spectral irregularities in the M87 X-ray spectrum. In this section,  we marginalise over the unknown magnetic field profile to derive constraints on the ALP parameters.
For a discussion of parameter constraints obtained using the  profile likelihood method, a Bayesian model comparison, and a discussion on the sensitivity of the bound to modelling assumptions and data reduction uncertainties, we refer the reader to Appendix \ref{app:A}. 


\subsection{Method}
\label{sec:method}
We first note that the observed spectrum of the AGN is well-fitted by an absorbed power-law (giving  $\chi^2 = 597.4$ for 614 degrees of freedom) in the 2--7 keV range,  and exhibits no large, irregular features that immediately call for physics beyond the Standard Model. 

To simulate the effects of an ALP on the observed AGN spectrum, we compute the survival probability function $P_{\gamma \to \gamma}(E)$ for photons propagating from M87 at the centre of the Virgo cluster to the virial radius, $R_{\rm vir} = 1.08\, {\rm Mpc}$. The survival probability is a sensitive function of the ALP parameters and the randomly generated magnetic field: we here sample 270 parameter choices in the relevant region of the $(m_a, \g)$ parameter space, and,  for each of these choices, compute $P_{\gamma \to \gamma}(E)$ for 100 randomly generated magnetic field profiles. 
Before entering the bulk of the Virgo cluster,
the soft X-ray energy range ($<2\, {\rm keV}$) of the power-law spectrum from the AGN gets modified
  by absorption in dense clouds near the black hole, resulting in an absorbed power-law spectrum that we denote by $I_0(E)$. The ALP-modified spectrum in the detector frame is given by, 
\be
I(E; b, m_a, \g) = 
D^{\rm MW}(E) P_{\gamma \to \gamma}(E(1+z); b, m_a, \g)  I_0(E(1+z))
\, ,
\ee 
where $b$ labels the random magnetic field profile  and $D^{\rm MW}(E)$  denotes the  absorption in the Milky Way. Here $z$ denotes the redshift to M87, \textbf{$z=0.0044$}. 

We used \textsc{flx2tab} in \textsc{xspec} to generate table models for each version of the ALP component.  The absorbed power-law model was then multiplied by a redshifted ALP table model and this combined model, $I(E; b, m_a, \g)$, was then fitted to the observed spectrum. Our subsequent  analysis of the result relies solely on the computed chi-square of each fit, $\chi^2(b, m_a, \g)$.

\subsection{Results}
\label{sec:constr2}
To obtain a constraint on the ALP parameters $(m_a, \g)$, we marginalise over the unknown magnetic field profile, as parametrised by $b$.
More precisely, under the assumption of approximate Gaussianity, we find the three-dimensional probability distribution $p(m_a, \g, b)$, which is normalised so that $\sum_{m_a,\, \g,\, b} p(m_a, \g, b) =1$. The marginalised probability distribution for the ALP parameters is then given by,
\be
p(m_a, \g) = \sum_b p(m_a, \g, b) \, .
\ee 

\begin{figure}
\centering
\includegraphics[width=0.95\textwidth]{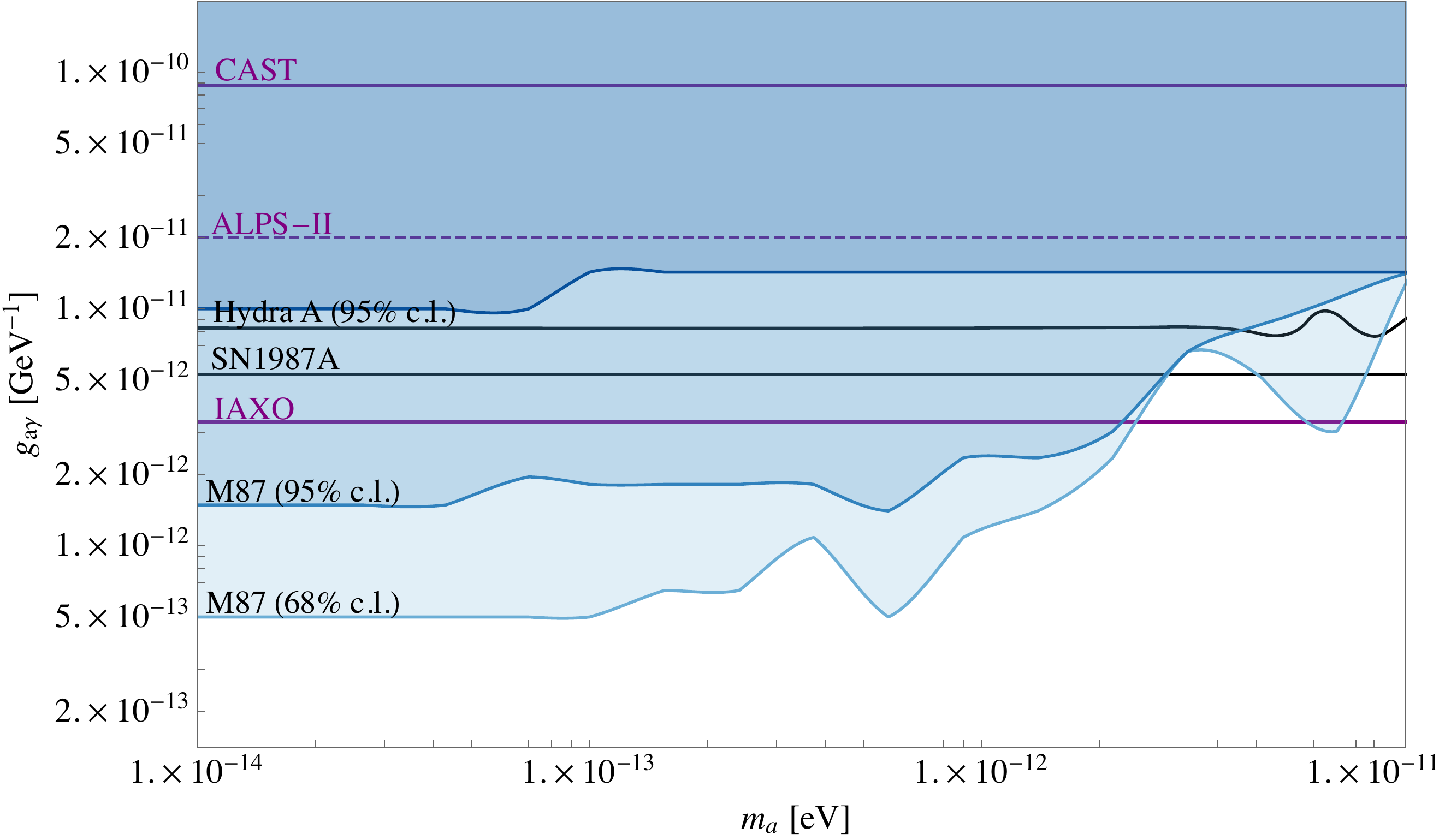}
\caption{Parameter constraints at the 68\%  (light blue), 95\%  (middle blue) and $99.7\%$ (dark blue) confidence levels. Coloured regions are excluded.
Not plotted are  constraints from NGC1275 which give $\g \lesssim 3.8$--$5.9\times10^{-12}\, {\rm GeV}^{-1}$ for $m_a \lesssim 10^{-12}\, {\rm eV}^{-1}$  \cite{Berg:2016ese}. 
}
\label{fig:Bayesian}
\end{figure}
Figure \ref{fig:Bayesian}, which is the main result of this paper, shows the inferred
 68\% and 95\% confidence regions in the ALP parameter space. We note that for light ALPs,  the upper bound on the ALP-photon coupling derived here,
 \be
 \g\big|_{B_0=31.5{\rm \mu G}} < 1.49 \times 10^{-12}\, {\rm GeV}^{-1} \, ,
 \ee 
  at 95\% confidence level,  is a factor of $\gtrsim 3$ stronger than the upper bound obtained from the analysis of SN987A  \cite{Brockway:1996yr, Grifols:1996id, Payez:2014xsa}. We note that 
   the transfer matrix only depends on the central magnetic field $B_0$ and $\g$ through $\Delta_{\gamma ai}$, and that the transformation $B_0 \to \lambda B_0$, $\g \to \g/\lambda$ leaves the conversion probability invariant. Our benchmark constraint  is derived for a conservative value of the central magnetic field ($B_0 = 31.5\, {\rm \mu G}$); should the actual magnetic field be stronger, our constraints would similarly be strengthened.
 %
 %
 %
 %
 For example, for a central magnetic field of  
 $B_0 = 40\, {\rm \mu G}$
 (as suggested by \cite{Owen}), keeping all else fixed, our constraint translates into,  
 \be
 \g \big|_{B_0=40{\rm \mu G}}  < 1.17 \times 10^{-12}\, {\rm GeV}^{-1} \, ,
 \ee
  at 95\% confidence level.  As shown in Appendix \ref{app:A}, these constraints are rather robust under modifications in the modelling of the magnetic field and the reduction of the observational data to take into account the cluster background. 


\section{Conclusions}
\label{sec:concl}
In this paper we have used \textit{Chandra} observations of the central point source of M87 in the Virgo cluster to search for  light ALPs. 
X-ray photons emitted from the AGN may oscillate into ALPs in the cluster magnetic field. As the oscillation probability  is energy dependent, this induces quasi-oscillatory features in the observed spectrum.  

We consider 25 {\it Chandra} observations with a total clean exposure time of $371.5\,{\rm ks}$, taken from 2010 to 2016. 
By only using observations with a short, $0.4\, {\rm s}$, frame time, the effects of pile-up are minimal and the cleaned data is of high quality. We analyse the nuclear spectra using \textsc{xspec} and find that it is well fitted by an absorbed power law, accounting for the flux variability, and exhibits no substantial modulations that call for Beyond the Standard Model physics. 

We construct models of the electron density and the magnetic field in the Virgo cluster that are consistent with X-ray and radio observations, and we use these models to 
simulate the effects of ALP-photon conversion from M87. As Virgo is a large cluster with a comparatively strong magnetic field, ALP-photon conversion is unsuppressed in a large region of the observationally allowed parameter space. 
We translate the absence of large modulations in the X-ray spectrum into constraints on the ALP parameters $(m_a, \g)$. 
Our new uppper bound on the ALP-photon coupling \g~is stronger than the bound from SN1987A for ALPs with masses $m_a < 3\times10^{-12}\, {\rm eV}$. 
Moreover, while our  method  is similar to that used to derive  
the bound from Hydra A \cite{Wouters:2013hua}, the high quality of the M87 data  leads to a substantially stronger bound on $\g$.

Interestingly, the values of \g~probed by M87 are in the range of future axion experiments, such as IAXO \cite{IAXO} and ALPS II \cite{ALPSIIa, ALPSIIb}, which, moreover, will be sensitive  to a much wider range of ALP masses, $m_a \lesssim 10^{-2}$--$10^{-4}\, {\rm eV}$. 

Our constraint is also interesting in relation to the suggested ALP explanation of the apparent anomalous transparency of the universe for very high energy cosmic rays \cite{Csaki:2003ef,Fairbairn:2009zi,Meyer:2013pny,Galanti:2015rda, Meyer:2014epa}, which postulate the existence of a light ALP with $m_a \lesssim 10^{-9}\, {\rm eV}$ and  $\g \gtrsim 3\times 10^{-12}\, {\rm GeV^{-1}}$.  Our bound constrains, but does not rule out, other possible hints of light ALPs such as the soft X-ray excess from galaxy cluster which may be explained by ALP-photon conversion of a relativistic background of ALPs \cite{Conlon:2013isa,Conlon:2013txa, Angus:2013sua, Marsh:2014gca,Powell:2014mda,Evoli:2016zhj}, and the explanation of the possible unidentified 3.5 keV line from galaxy clusters \cite{Bulbul:2014sua, Boyarsky:2014jta}, which may be explained by dark matter decaying into ALPs, which subsequently convert to photons \cite{Cicoli:2014bfa, Alvarez:2014gua,Conlon:2014wna}.

Finally, we note that the constraints derived in this paper also apply to other scalar particles with a dilaton-like (or Chameleon-like), $\phi\, F_{\mu \nu} F^{\mu \nu}$, coupling to electromagnetism. Our bound is two to three orders of magnitude stronger than earlier constraints on Chamelon-like particles from galaxy clusters  \cite{Davis:2010nj} (see also \cite{Wouters:2013hua}).  

In conclusion, our result highlights the power of X-ray astronomy in searches for axion-like particles. We expect that complementary constraints can be obtained from other localised, bright X-ray sources, and possibly also from the thermal spectrum of cluster gas. 

\vspace*{1 cm}
\subsection*{Acknowledgements}
We would like to thank Ben Allanach and Mark Manera for stimulating discussions, and Joseph Conlon for discussions and valuable comments on the draft.  
 DM is supported by a Stephen Hawking Advanced Fellowship at the Centre for Theoretical Cosmology, University of Cambridge. 
HRR and ACF acknowledges support from ERC Advanced Grant Feedback 340442. 
CSR thanks support from the Chandra Guest Observer Program under the Smithsonian Astrophysical Observatory grant GO617110B. Support for this work was provided by the National Aeronautics and Space Administration through Chandra Award Number GO6-17110A issued by the Chandra X-ray Observatory Center, which is operated by the Smithsonian Astrophysical Observatory for and on behalf of the National Aeronautics Space Administration under contract NAS8-03060.

\appendix
\section{Sensitivity to background subtraction and modelling assumptions}
\label{app:A}
In this appendix, 
 we 
show that a naive application of the  frequentist's profile likelihood method leads to a strong  hint of a light ALP at $99\%$ confidence level from M87, but that this 
hint is spurious and  due to overfitting statistical fluctuations.
Moreover,
we investigate the sensitivity of our constraint to the assumptions of the data reduction and the cluster modelling. 

\subsection{Profile likelihood constraints and over-fitting}
\label{sec:constr1}
 For particular values of the parameters $(m_a, \g)$, the ALP-induced modulations are comparable in size to the statistical fluctuations of the binned X-ray data. 
For certain magnetic fields, the  
ALP-modulated AGN spectrum can then lead to a better fit than the plain absorbed power-law by 
over-fitting statistical fluctuations. This effect was found in a  likelihood  analysis of X-ray data from Hydra A  to lead to a slight ($1.2\sigma$) statistical preference for a non-vanishing ALP coupling \cite{Wouters:2013hua}, which, however, should not be interpreted as evidence for a new particle. In this section, we  show that this effect becomes more pronounced with improved observational data, leading in our case to an apparent exclusion of a vanishing ALP-photon coupling at the 99\% confidence level.  
We note  that this limits the naive applicability of standard frequentist's tools for constraining ALP parameters from astrophysical observations.

\begin{figure*}
\centering
\includegraphics[width=0.55\textwidth]{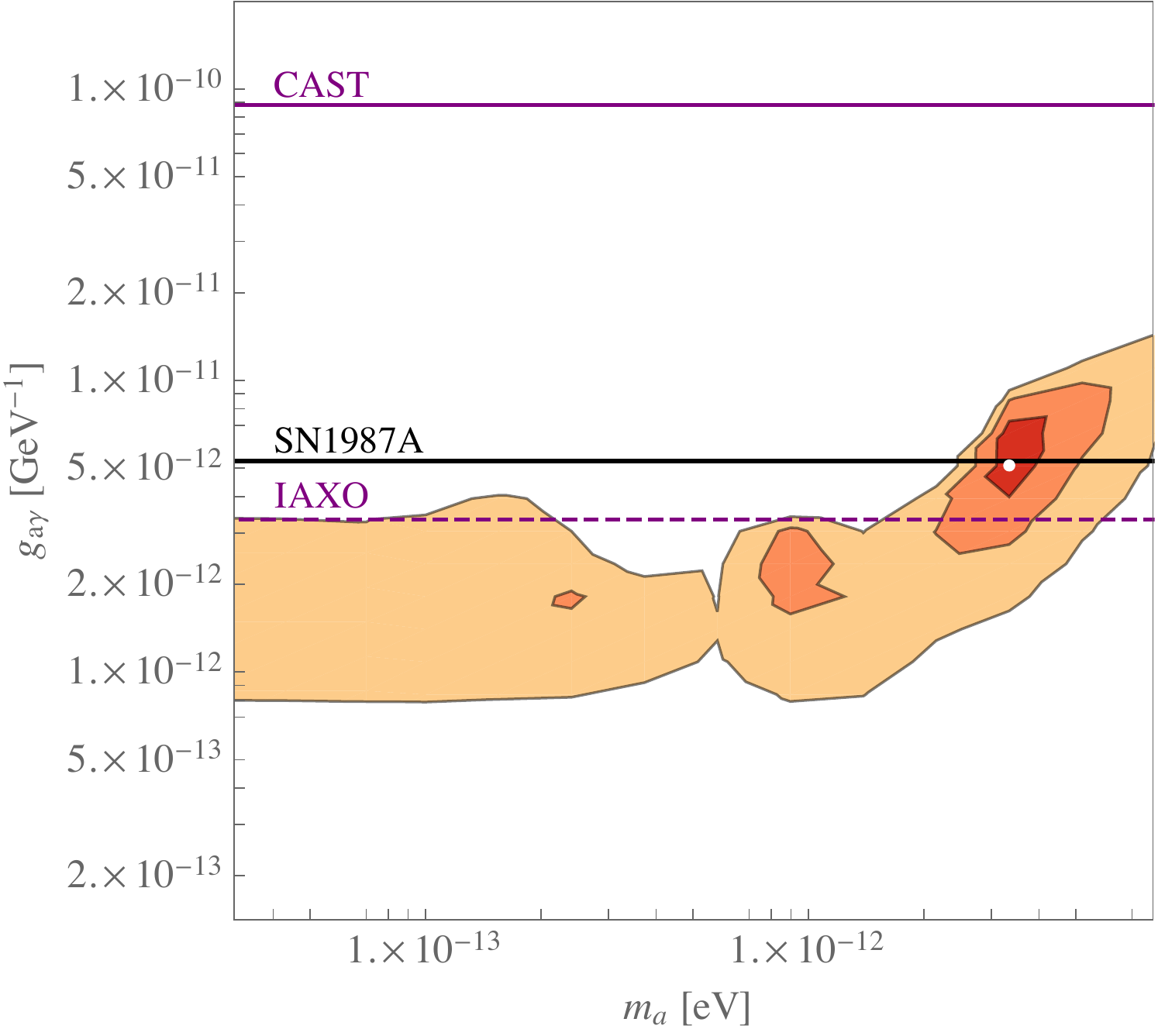}
\caption{Profile likelihood parameter constraints at 68\% (red), 95\% (orange) and 99\% (yellow) confidence levels; the white dot indicates the best-fit parameters.   }
\label{fig:constr}
\end{figure*}

\begin{figure*}
\centering
\includegraphics[width=0.5\textwidth]{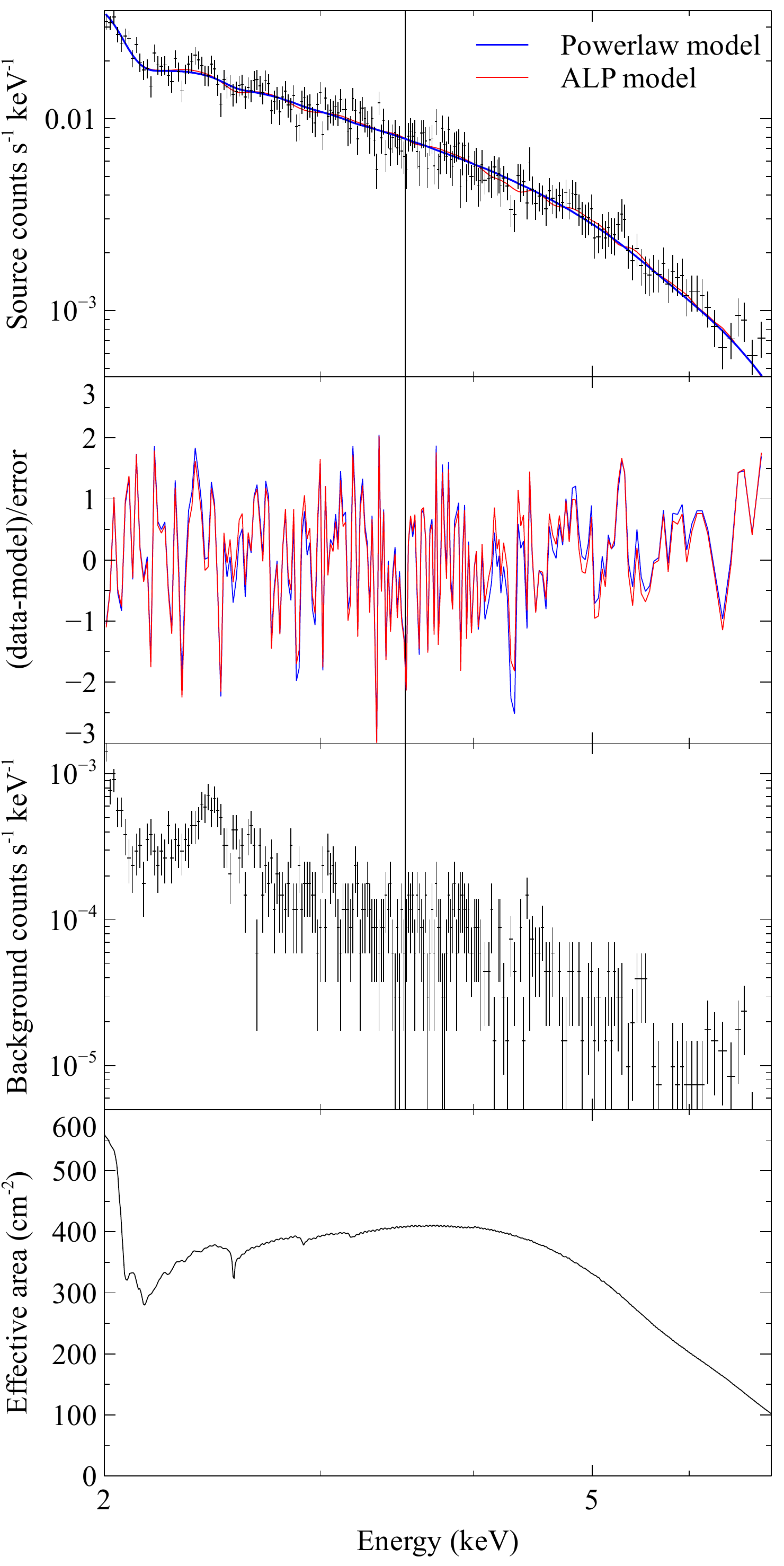}
\caption{Comparison of the best-fit ALP model with the best-fit power-law model.  The panels show the summed 2016 spectrum (see Fig. \ref{fig:spectra}) with the two best-fit models, the residuals for each model, the cluster background spectrum and \textit{Chandra's} effective area.
}
\label{fig:bestfit}
\end{figure*}

\begin{figure*}
\centering
\begin{minipage}{\textwidth}
\includegraphics[width=0.45\textwidth]{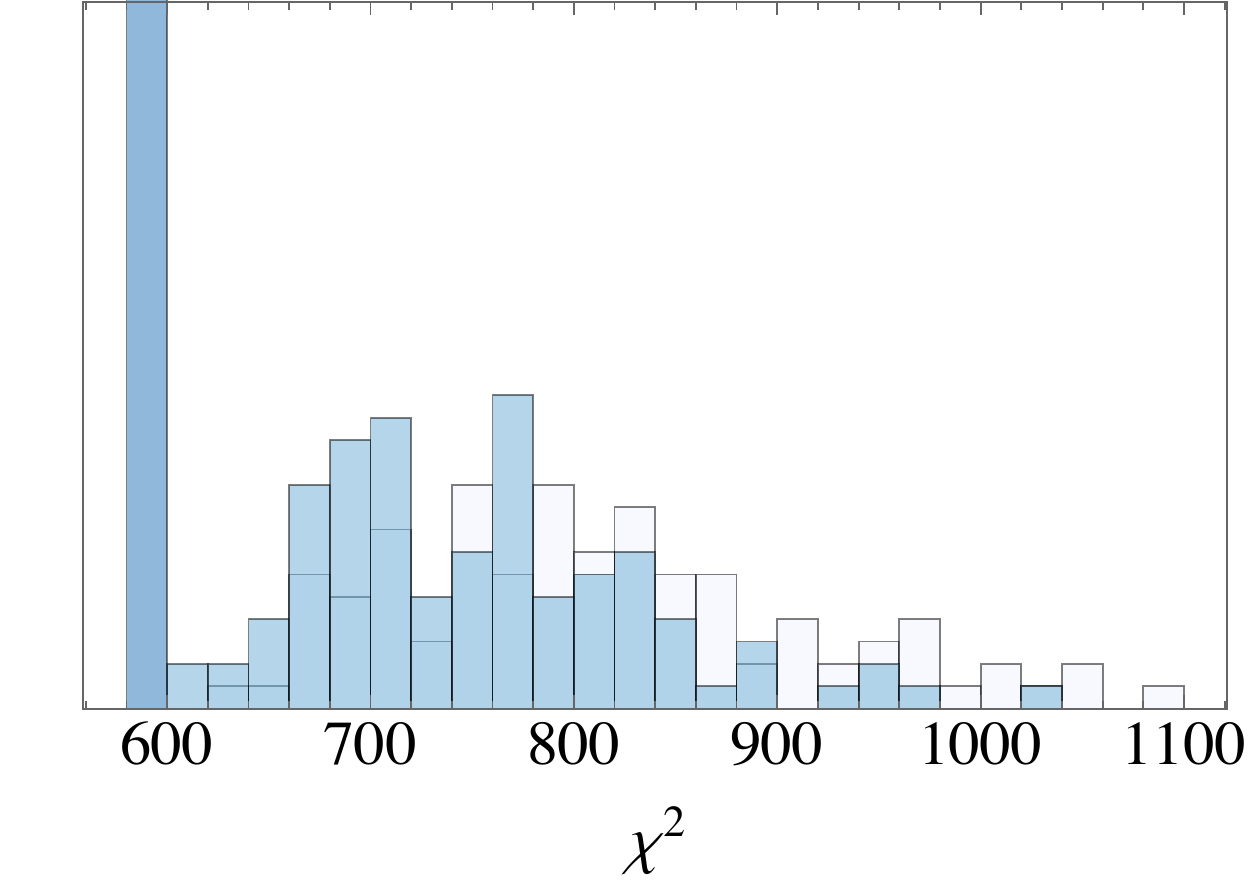}
\includegraphics[width=0.45\textwidth]{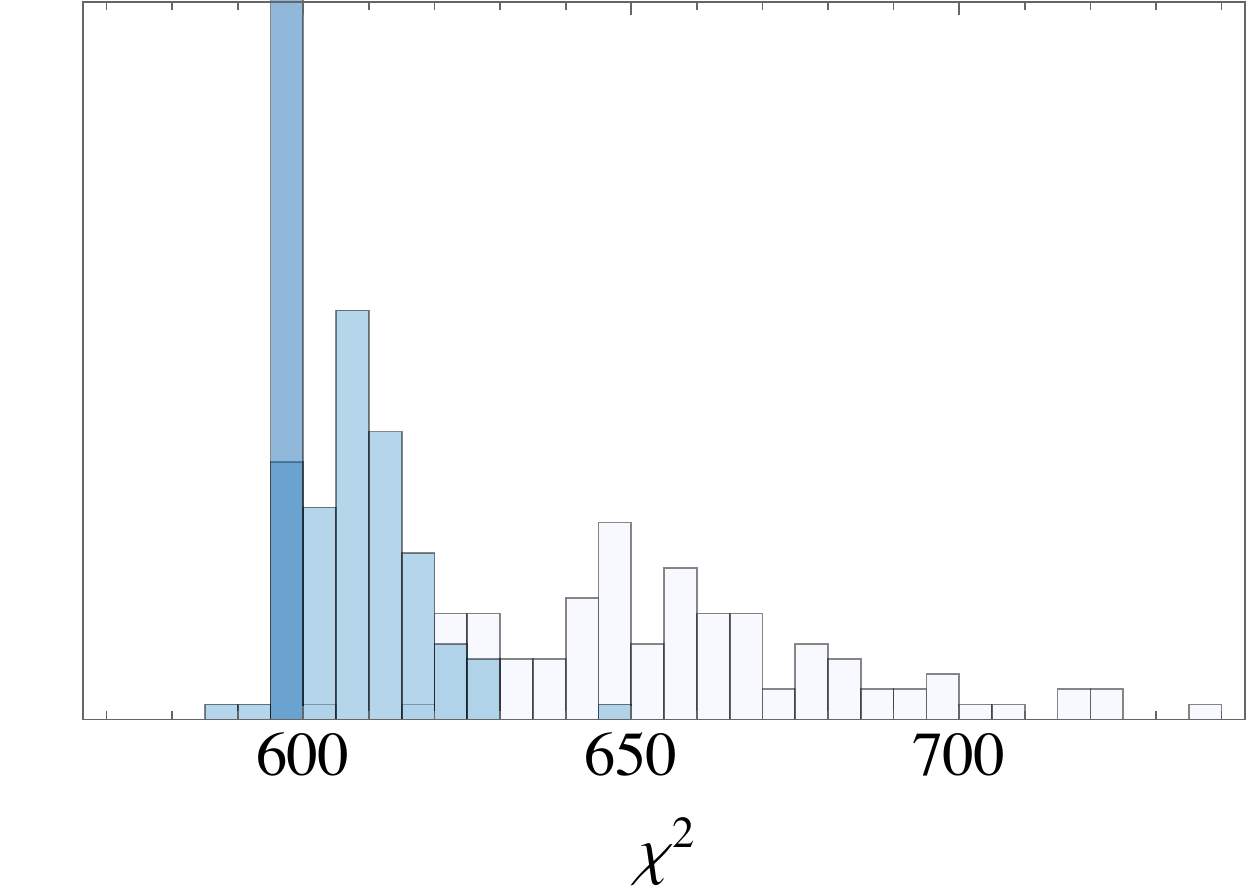}
\caption{Distribution of $\chi^2$ parameters for 100 realisations of the cluster magnetic field. Here, $m_a=0$ (left) and $m_a=3.3\times10^{-12}\, {\rm eV}$ (right); $\g= 1.1\times10^{-11}\, {\rm GeV^{-1}}$ (whitish), $\g=5.1\times10^{-12} \, {\rm GeV^{-1}}$ (light blue) and  $\g=5\times10^{-13} \, {\rm GeV^{-1}}$ (darker blue). The area under each curve is the same and the dark blue column is vertically truncated for the sake of presentation.  
}
 \label{fig:chisq}
\end{minipage}
\end{figure*}

Assuming the existence of an ALP, its parameters can be constrained  using the profile likelihood method. We here follow the prescription of
\cite{Avni}, which we  now review. 
We denote  the X-ray observations $O_i$ (for $i=1, \ldots, n$) and  the inverse covariance matrix of these measurements  by $\Sigma^{ij}$. The data is modelled by the values  $C_i(\Theta)$ that depend on the model parameters $\Theta$. The best-fit values of the parameters are obtained by minimising the $\chi^2$ function,
\be
\chi^2(\Theta) = \left( O_i - C_i(\Theta) \right) \Sigma^{ij} \left( O_j - C_j(\Theta) \right) \, ,
\ee
where repeated indices are summed over. The model parameters $\Theta$ include the `interesting' axion parameters $\Phi = \{m_a, \g\}$ as well as the (for our purposes) uninteresting parameters $\Psi$: these include the spectral index of the source, $n_s$, the hydrogen column densities for each of the observations, $n_H^{({\rm obs})}$, and the magnetic field realisation $b$. For a given confidence level $\alpha$, the confidence region $R(\alpha)$ for $\Phi$ is determined by  the axion parameters satisfying,
\be
\underset{\Psi}{\rm min}\, \left(\chi^2( \Phi, \Psi)\right) - \underset{\Psi, \, \Phi}{\rm min}\left(\chi^2(\Phi, \Psi) \right)\leq \Delta(\alpha) \, ,
\label{eq:Ralpha}
\ee
where $\Delta(\alpha)$ is determined by,
\be
{\rm Probability}\left(\chi^2_{{\rm dim}(\Phi)\, {\rm d.o.f.}} \leq \Delta(\alpha)\right) \leq \alpha \, .
\ee
The results of this analysis are presented  in Figure \ref{fig:constr}. The best-fit parameters are given by $m_a = 3.3\times10^{-12}\, {\rm eV}$ and $\g = 5.1 \times10^{-12}\, {\rm GeV}^{-1}$, and are located just below the bound from the absence of an associated gamma-ray burst from SN1987A \cite{Payez:2014xsa}. The survival probability $P_{\gamma\to\gamma}(E)$ of the best-fit model is given by the black curve of Figure \ref{fig:Pexpl}. The 68\% and 95\% confidence level contours rather narrowly encircle the best-fit point, and even the 99\% contour excludes a vanishing ALP-photon coupling, ostensibly leading to a `hint' of an axion-like particle. However, this hint is spurious: the minimisation prescription of equation \eqref{eq:Ralpha} selects the magnetic field profile that produces the most well-fitted modulation pattern, but the actual Virgo magnetic field 
is unknowable by any existing means, 
and may be quite different from that selected by this method.  
A frequentist analysis that properly takes the `look elsewhere' effect into account should drastically decrease the significance of this hint, and provide constraints  consistent with the Bayesian analysis in section \ref{sec:constr2}.
Figure \ref{fig:bestfit} shows the best-fit ALP model together with  the unmodulated power-law model. While the unmodulated power spectrum has $\chi^2 =597$ for 614 degrees of freedom, the best-fit ALP model has  $\chi^2=586$ for 617 degrees of freedom.
The distribution of $\chi^2$ values for  different choices of the ALP parameters $m_a$ and $\g$  are shown in Figure \ref{fig:chisq}. Finally, 
Figure \ref{fig:altFits} shows two additional ALP-distorted spectra and
Figure \ref{fig:PofE_multiple} shows $P_{\gamma\to a}(E)$ for four random realisation of the cluster magnetic field, and nine choices of the ALP parameters.  

\subsection{Bayesian model comparison}
 One way to compare how well the data fits the ALP model, which we will denote ${\cal M}_1$, and the unmodulated, no-ALP model, ${\cal M}_0$, is to compute the corresponding Bayes factor (for a recent review of Bayesian statistics, see e.g.~\cite{Trotta}).   
 The data provides increased evidence for model ${\cal M}_1$ if the Bayes factor satisfies,
\be
B_{01} \equiv \frac{p(d|{\cal M}_0)}{p(d|{\cal M}_1)}  < 1 \, .
\ee
Here, $p(d|{\cal M})$ denotes the Bayesian evidence,
\be
p(d|{\cal M}) \equiv \int p(d| \Theta, {\cal M}) p(\Theta|{\cal M}) d \Theta \, ,
\label{eq:BayesEv}
\ee  
where $ p(\Theta|{\cal M})$ denotes the prior probability distribution of the model parameters.
 For the ALP model, we have $\Theta = \{m_a, \g, b \}$, while the no-ALP model has no free parameters.\footnote{ For simplicity, we exclude the hydrogen column density and the source spectral index from the free parameters of the model. These parameters have the same parameter space for both ${\cal M}_0$ and ${\cal M}_1$, and are unlikely to affect the conclusions of this section.} Recognising our ignorance of the detailed magnetic field profile of the Virgo cluster, we take a uniform prior distribution for $b$. The priors of  $m_a$ and \g~are chosen to be uniform  on a log scale. We furthermore  approximate the model parameters as Gaussian variables, so that  the likelihoods scale like $\sim {\rm exp}(-\chi^2/2)$.  We then find,
\be
B_{01} = 2.0 \, ,
\ee 
which corresponds to a small increase in evidence for the no-ALP model, ${\cal M}_0$. When interpreted, as is commonly done, against the empirical Jeffrey's scale, this amounts to inconclusive evidence.

In sum, we have shown that a naive application of the profile likelihood method leads to a rather strong hint for a non-vanishing ALP coupling, but that both direct investigation of the corresponding best-fit model and a Bayesian model comparison point towards this hint being spurious. 

 \subsection{Magnetic field modelling sensitivity}
\begin{figure*}
\centering
\begin{minipage}{\textwidth}
\includegraphics[width=0.48\textwidth]{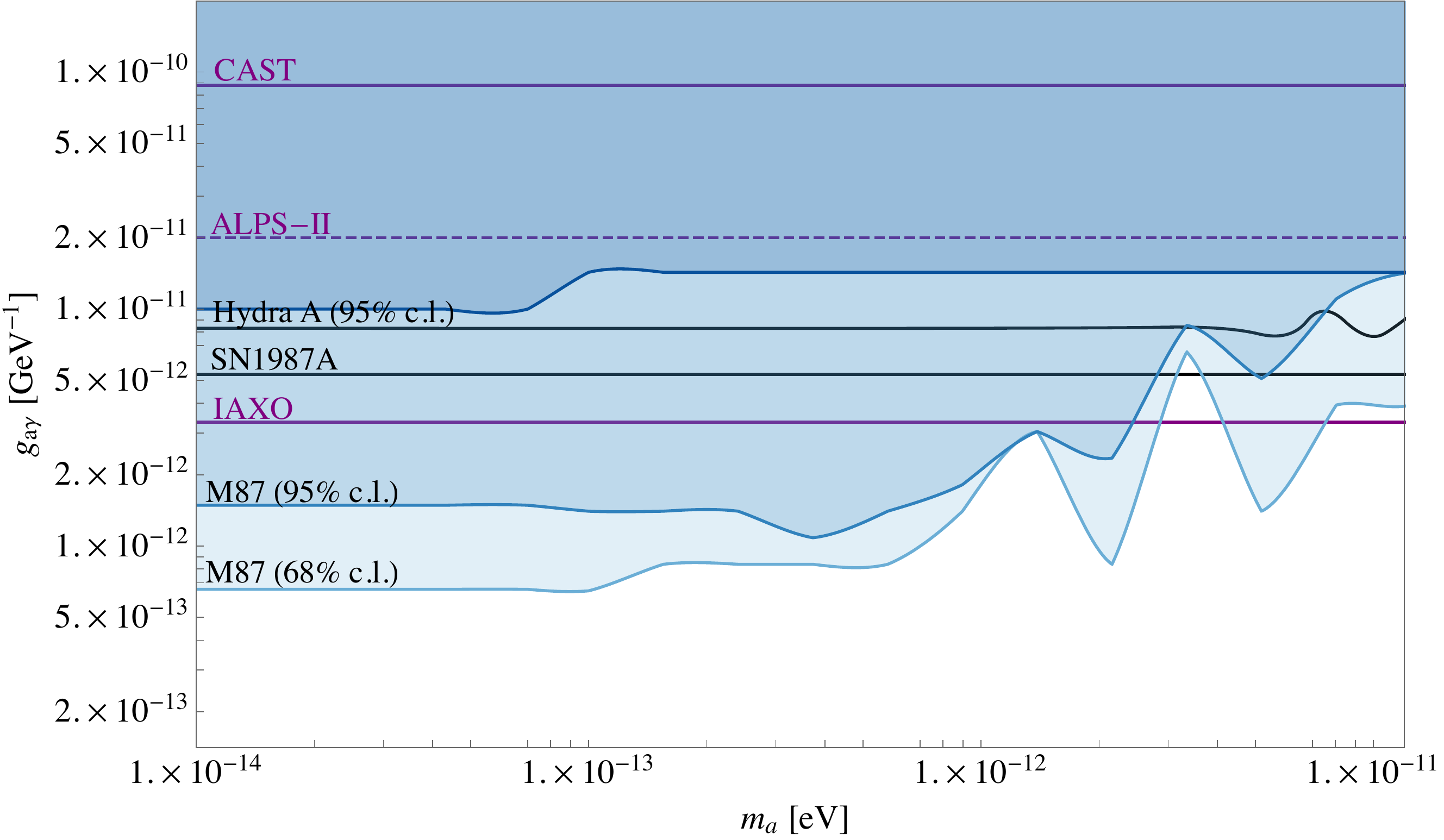}
\includegraphics[width=0.48\textwidth]{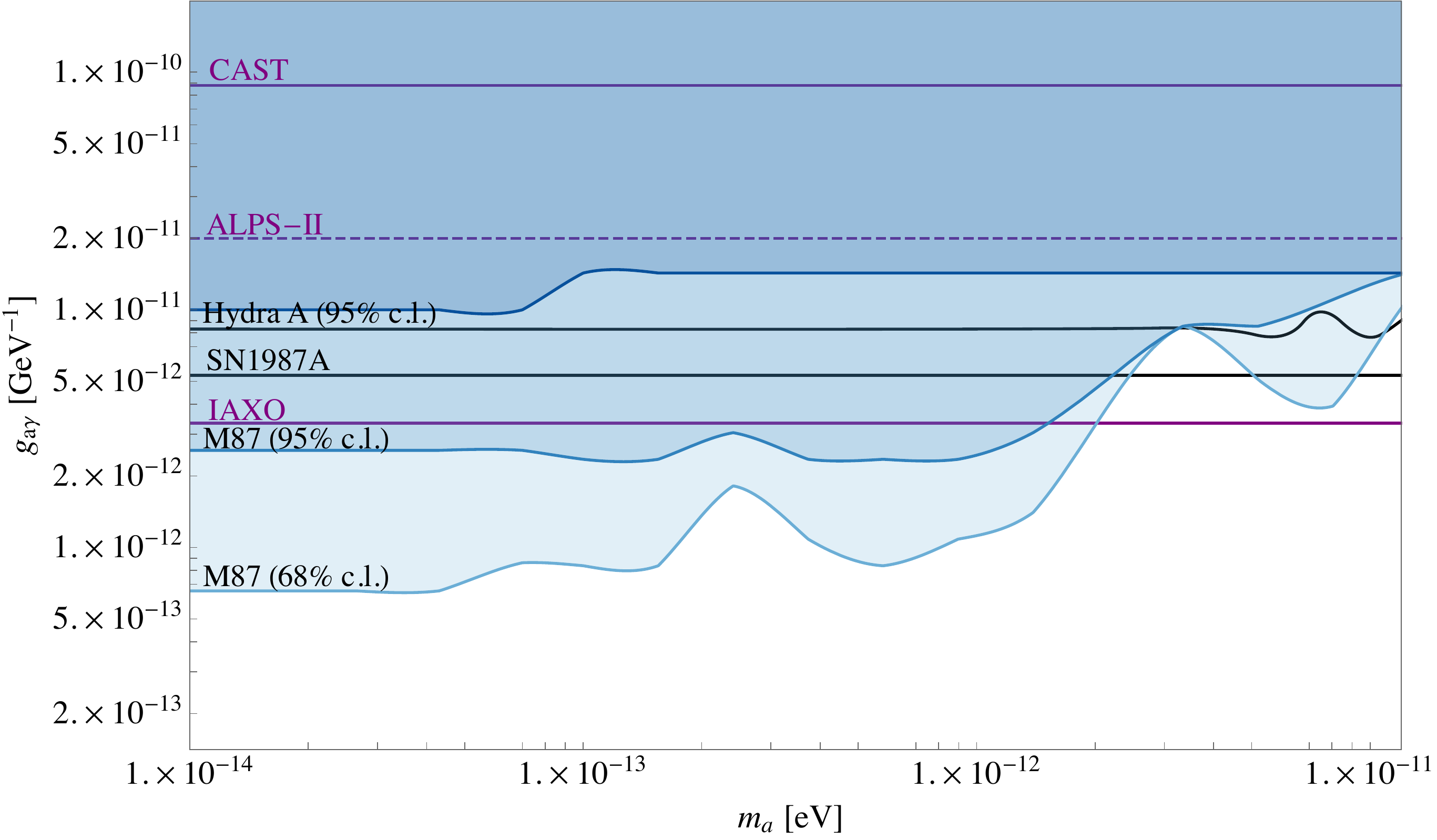}
\caption{Parameter constraints for alternative magnetic field model 1 (left) and 2 (right)
 at the 68\%  (light blue), 95\%  (middle blue) and $99.7\%$ (dark blue) confidence levels.
Coloured regions are excluded.}
\label{fig:altB}
\end{minipage}
\end{figure*}
Radio observations of M87 and M84 constrain the magnetic field of the Virgo cluster,
but they do not provide a unique solution for the magnetic field model parameters. In this section, we investigate the robustness of the constraint on the ALP parameters under certain changes in the magnetic field model. 

The stochastic magnetic field model of section \ref{sec:Bmodel} has four parameters which for the benchmark model are given by,
\be
(B_0, \gamma, \alpha, L_{\rm min}) 
=
(31.6\, {\rm \mu G}, 2.5, 0.54, 1\, {\rm kpc})
\, .
\ee 
The dependence on the central magnetic field $B_0$ is trivial: the transfer matrix has a symmetry $B_0 \to \lambda B_0$, $\g \to \g/\lambda$, and a strengthened central magnetic field corresponds to a similarly strengthened upper bound on \g. A smaller value of $\gamma$ corresponds to fewer small and more large 
domains, which generically increases the conversion probability. Similarly, a larger value of $L_{\rm min}$ increases the ALP-photon conversion probability. Finally, a larger value of $\alpha$ increases the rate at which the magnetic fields falls of with radius, which suppresses the conversion probability. 
In section \ref{sec:Bmodel} we constrained the possible values of $B_0$ and $\alpha$ by Faraday RMs from M87 and M84. Observations of additional radio sources in or behind the Virgo cluster would further constrain the magnetic field model.


To investigate the sensitivity to the magnetic field parameters, we repeat the full analysis for two additional magnetic field models. For the first alternative magnetic field model we take $L_{\rm min} = 3.5\, {\rm kpc}$, with all other parameters as in the benchmark magnetic field model.  For the second alternative magnetic field model, we take $B_0 = 35\, {\rm \mu G}$, $L_{\rm min} = 1\, {\rm kpc}$, and chose a steep magnetic field fall-off, $\alpha = 0.67$. As we noted in section \ref{sec:Bmodel}, the benchmark value of $\alpha \approx 1/2$ can be motivated by matching the scaling of the energy density in the magnetic field to that of the gas. This value is also consistent with more elaborate magnetic field models from other clusters \cite{A665}. However, the value of $\alpha=2/3$ may arise if the magnetic field is  `frozen into' the plasma, and is not excluded by Faraday rotation observations.

The corresponding parameter constraints are presented in Figure \ref{fig:altB}. The constraints from both alternative models are in good qualitative agreement with those derived from the benchmark magnetic field model, and for  alternative model 1, even the  upper bound on \g~for very light ALPs is unchanged. For model 2, the steep fall-off of the magnetic field results in a slightly 
higher 95\% contour, while the 68\% contour remains essentially unchanged. 
%
%
Hence, we note that our constraint changes by less than a factor of two under modifications of the underlying magnetic field model, and the upper bound on \g~is in all cases stronger than that derived from SN1987A.

\subsection{Cluster background subtraction}
\label{sec:bkgrnd}
\begin{figure*}
\centering
\includegraphics[width=0.65\textwidth]{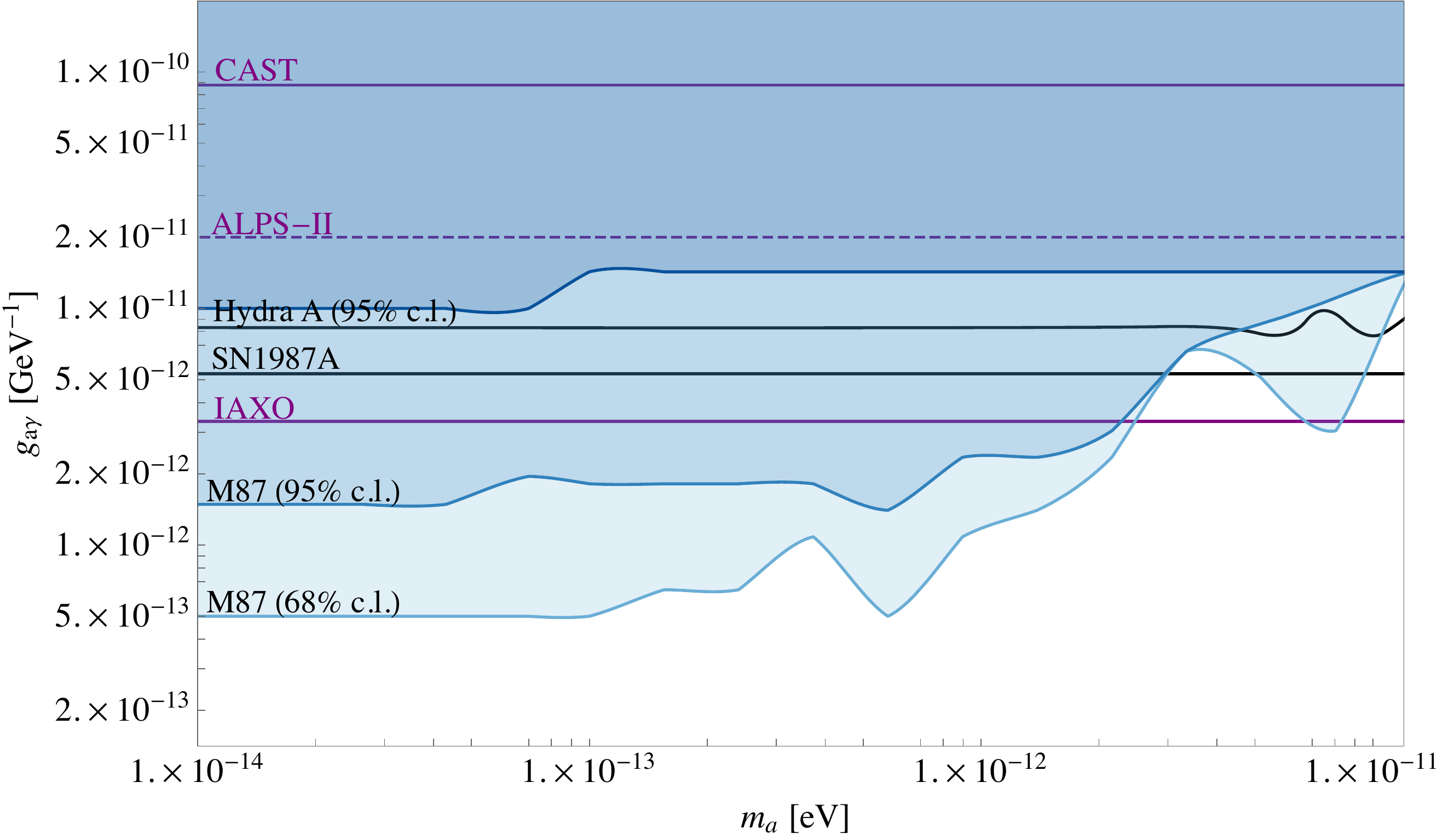}
\caption{Original magnetic field model, 15\% additional background subtraction.  
}
\label{fig:15pc}
\end{figure*}
To investigate the sensitivity of the constraints to the accuracy with which the cluster background is subtracted, we re-run the \textsc{xspec} optimisation using the observed nuclear spectrum and a 15\% brighter background spectrum.  Fig. \ref{fig:chartfigs} shows that, whilst the cluster surface brightness profile over the energy range $2-7\keV$ is approximately flat over the central few arcsec, it is also consistent with a modest increase of $\sim10\%$ within the errors. We therefore test an increase the cluster background spectrum by a conservative value of $15\%$. The results are presented in Figure \ref{fig:15pc} and show that this level of uncertainty in the cluster background has a negligible effect on the ALP parameter constraints.

\begin{figure*}
\centering
\includegraphics[width=\textwidth, height=350pt]{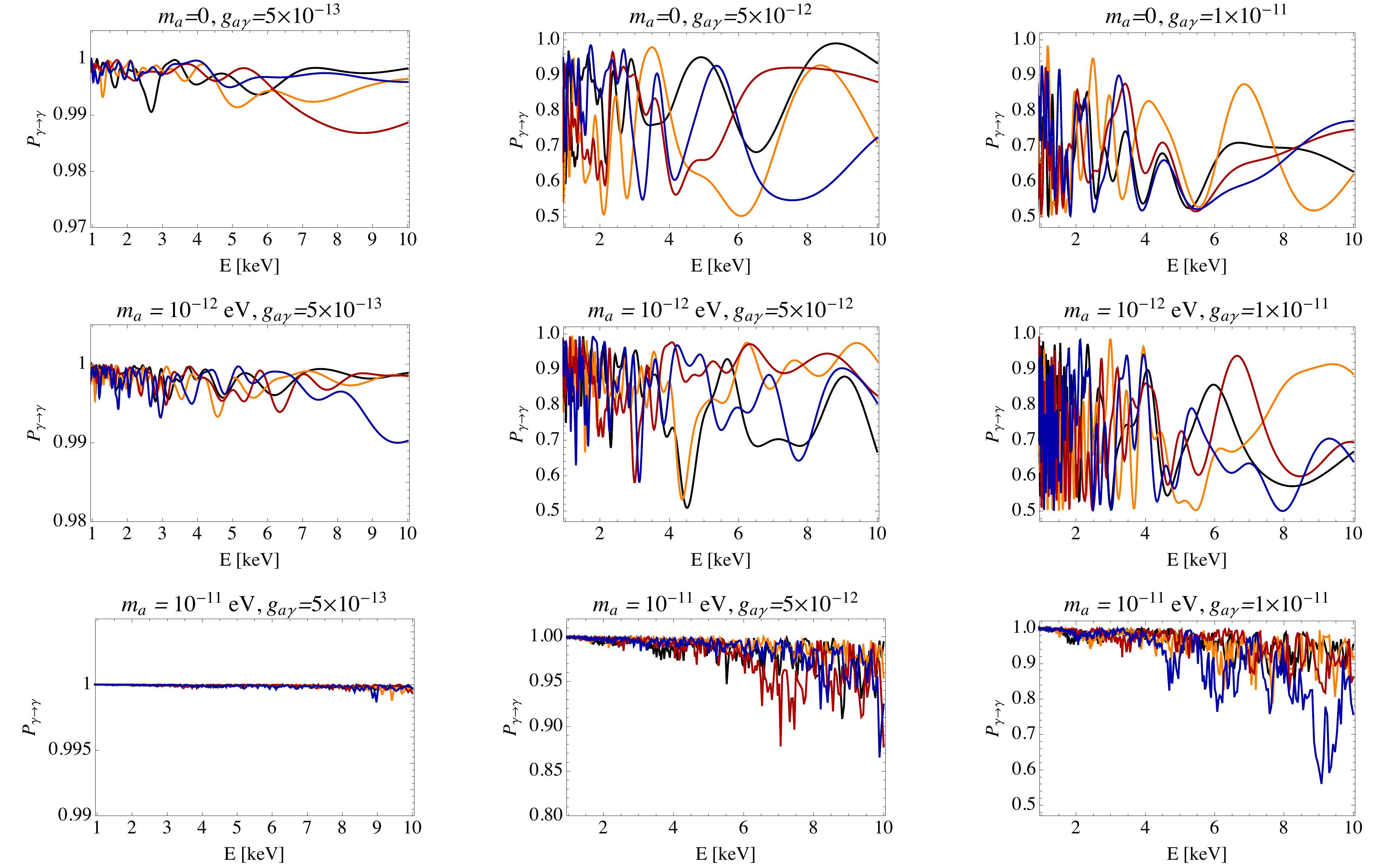}
\caption{Conversion probability as function of energy for four random realisations of the cluster magnetic field (blue, black, orange and red), and various choices of ALP parameters. }
\label{fig:PofE_multiple}
\end{figure*}

\begin{figure*}
\centering
\begin{minipage}{\textwidth}
\includegraphics[width=0.45\textwidth]{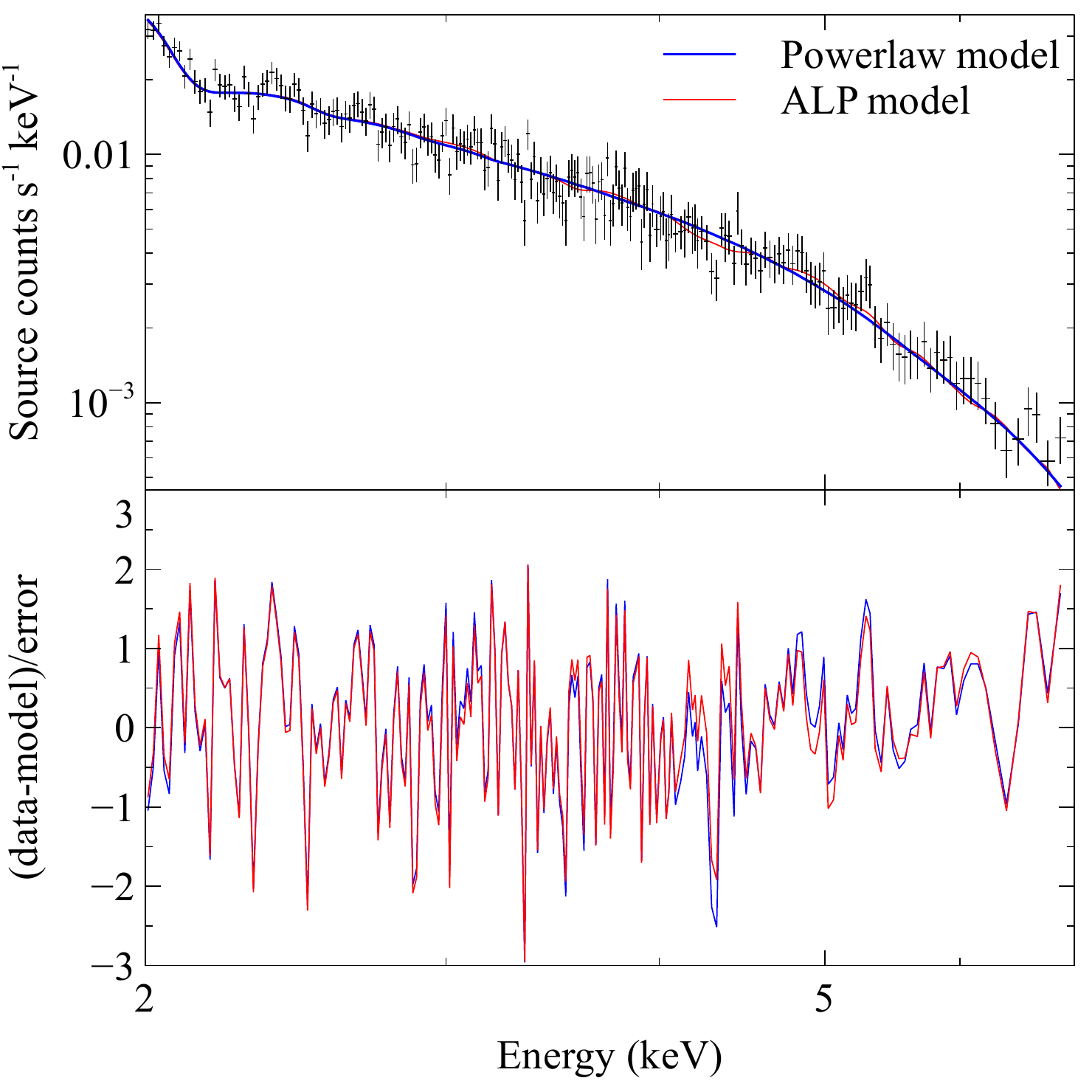}
\includegraphics[width=0.45\textwidth]{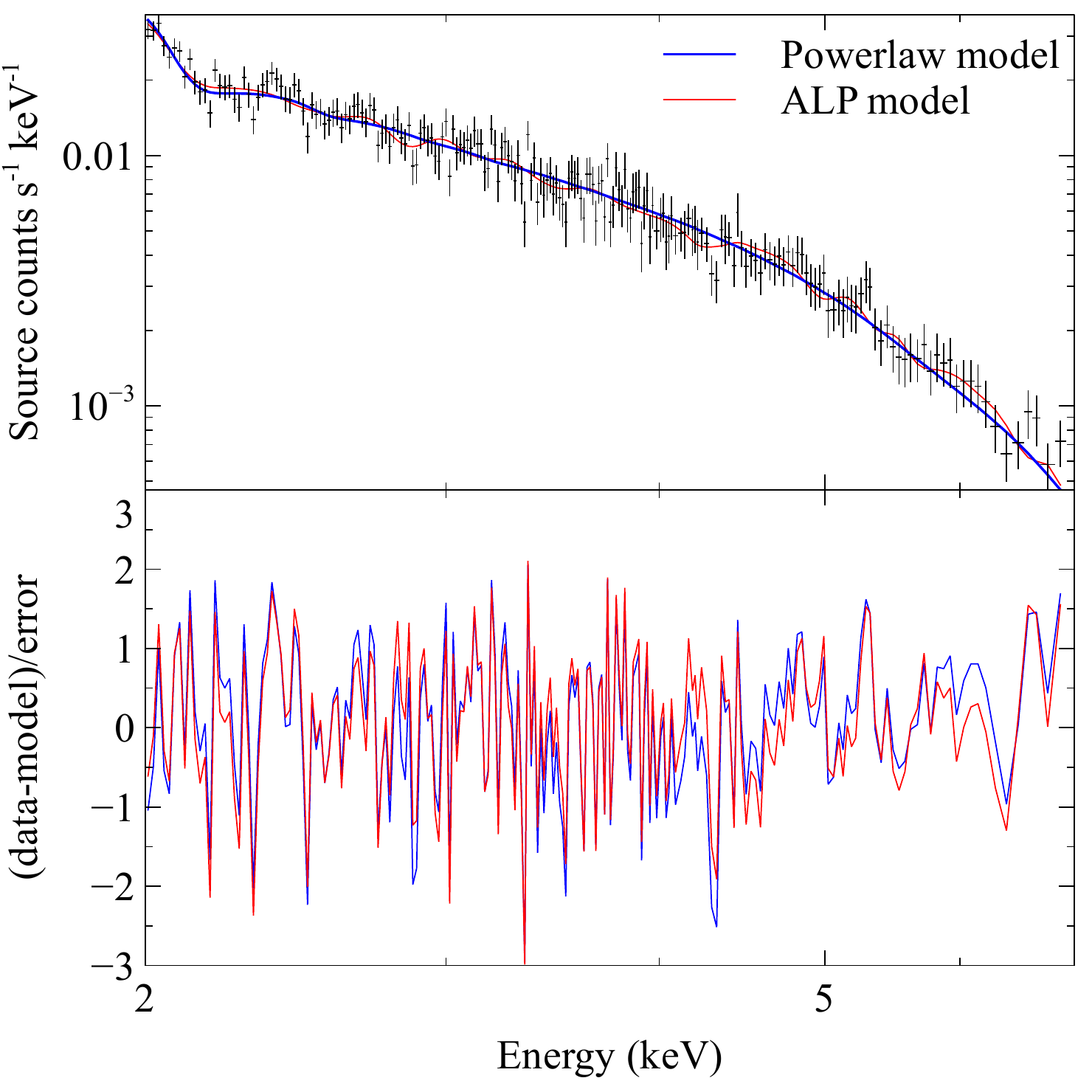}
\caption{Comparisons of two alternative ALP models with the best-fit power-law model. 
The left model has $m_a = 2\times 10^{-12}\, {\rm eV}$ and $\g= 3\times 10^{-12}\, {\rm GeV}^{-1}$ and gives a fit with $\chi^2 = 592$, falling within the 95\% contour of Figure \ref{fig:constr}. The right model has $m_a = 2\times 10^{-12}\, {\rm eV}$ and $\g= 5\times 10^{-12}\, {\rm GeV}^{-1}$ and gives a fit with $\chi^2 = 599$, falling just outside the 99\% contour.
}
 \label{fig:altFits}
\end{minipage}
\end{figure*}

\pagebreak

\bibliographystyle{JHEP}
\bibliography{references}

\end{document}